\documentclass[fleqn,usenatbib]{mnras}

\usepackage{newtxtext,newtxmath}

\usepackage[T1]{fontenc}

\DeclareRobustCommand{\VAN}[3]{#2}
\let\VANthebibliography\thebibliography
\def\thebibliography{\DeclareRobustCommand{\VAN}[3]{##3}\VANthebibliography}

\usepackage{graphicx}	
\usepackage{amsmath}	
\usepackage{afterpage}
\usepackage{lscape}
\usepackage{academicons}
\usepackage{upgreek}

\title[GC system of NGC\,4382]{The complex globular cluster system of the S0 galaxy NGC\,4382 in the outskirts of the Virgo Cluster}

\author[Escudero et al.]{Carlos G. Escudero,$^{\href{https://orcid.org/0000-0002-6056-6247}{\includegraphics[width=7pt]{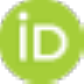}}\,1,2,3}$\thanks{E-mail: cgescudero@fcaglp.unlp.edu.ar}
Arianna Cortesi,$^{4}$
Favio R. Faifer,$^{1,2,3}$
Leandro A. Sesto,$^{1,2,3}$
\newauthor{Analía V. Smith Castelli,$^{1,2,3}$
Evelyn J. Johnston,$^{5}$
Victoria Reynaldi,$^{1}$
Ana L. Chies-Santos,$^{7,8}$}
\newauthor{Ricardo Salinas,$^{6}$
Kar\'in Men\'endez-Delmestre,$^{4}$
Thiago, S. Gon\c{c}alves,$^{4}$
Marco Grossi,$^{4}$}
\newauthor{Claudia Mendes de Oliveira,$^{9}$
}
\\
$^{1}$Facultad de Cs. Astron\'omicas y Geof\'isicas, UNLP, Paseo del Bosque S/N, 1900 La Plata, Argentina\\
$^{2}$Instituto de Astrof\'isica de La Plata (CCT La Plata - CONICET - UNLP)\\
$^{3}$Consejo Nacional de Investigaciones Cient\'ificas y T\'ecnicas, Rivadavia 1917, C1033AAJ Ciudad Aut\'onoma de Buenos Aires, Argentina\\
$^{4}$Federal University of Rio de Janeiro, Valongo Observatory, Ladeira Pedro Antonio, 43, Saude 20080-090 Rio de Janeiro, Brazil \\
$^{5}$N\'ucleo de Astronom\'ia de la Facultad de Ingenier\'ia y Ciencias, Universidad Diego Portales, Av. Ej\'ercito Libertador 441, Santiago, Chile \\
$^{6}$Gemini Observatory, Casilla 603, La Serena, Chile \\
$^{7}$ Instituto de F\'isica, Universidade Federal do Rio Grande do Sul (UFRGS), Av. Bento Gon\c{c}alves, 9500, Porto Alegre, RS, Brazil \\
$^{8}$ Shanghai Astronomical Observatory, Chinese Academy of Sciences, 80 Nandan Rd., Shanghai 200030, China \\
$^{9}$Departamento de Astronomia, Instituto de Astronomia, Geof\'isica e Ci\^encias Atmosf\'ericas da USP, Cidade Universit\'aria, 05508-900, S\~ao Paulo, SP, Brazil
}

\date{Accepted XXX. Received YYY; in original form ZZZ}

\pubyear{2015}

\begin{document}
\label{firstpage}
\pagerange{\pageref{firstpage}--\pageref{lastpage}}
\maketitle

\begin{abstract}
NGC\,4382 is a merger-remnant galaxy that has been classified as morphological type E2, S0, and even Sa. In this work, we performed a photometric and spectroscopic analysis of the globular cluster (GC) system of this peculiar galaxy in order to provide additional information about its history. We used a combination of photometric data in different filters, and multi-object and long-slit spectroscopic data obtained using the Gemini/GMOS instrument. The photometric analysis of the GC system, using the Gaussian Mixture Model algorithm in the colour plane, reveals a complex colour distribution within $R_\mathrm{gal}<5$ arcmin (26.1 kpc), showing four different groups: the typical blue and red subpopulations, a group with intermediate colours, and the fourth group towards even redder colours. From the spectroscopic analysis of 47 GCs, confirmed members of NGC\,4382 based on radial velocities, we verified 3 of the 4 photometric groups from the analysis of their stellar populations using the ULySS code. 
NGC\,4382 presents the classic blue ($10.4\pm2.8$ Gyr, $\mathrm{[Fe/H]}=-1.48\pm0.18$ dex) and red ($12.1\pm2.3$ Gyr, $\mathrm{[Fe/H]}=-0.64\pm0.26$ dex) GCs formed earlier in the lifetime of the galaxy, and a third group of young GCs ($2.2\pm0.9$ Gyr; $\mathrm{[Fe/H]}=-0.05\pm0.28$ dex). Finally, analysis of long-slit data of the galaxy reveals a luminosity-weighted mean age for the stellar population of $\sim$2.7 Gyr, and an increasing metallicity from [Fe/H]=$-0.1$ to $+0.2$ dex in $R_\mathrm{gal}<10$ arcsec (0.87 kpc). These values, and other morphological signatures in the galaxy, are in good agreement with the younger group of GCs, indicating a common origin as a result of a recent merger. 

\end{abstract}




\section{Introduction}
\label{intro}

Globular clusters (GCs) are key players in the evolutionary history of galaxies since they are tracers of their main star-formation episodes \citep[e.g.,][]{Blom2014,Sesto2018,Forte2019,Fensch2020}. Being objects more luminous than the starlight of the galaxy in which they are associated, GCs allow obtaining information from their host galaxy at large galactocentric radii \citep[e.g.,][]{Norris2008,Pota2013,Dolfi2020,Dolfi2021}.

GC systems in the most massive galaxies are known to present a bimodal optical colour distribution \citep[e.g.,][]{Elson1996,Larsen2001,Faifer2011,Escudero2018}. This bimodality in colour is translated into bimodal metallicity distributions \citep[e.g.,][]{Cenarro2007,Brodie2012,Usher2012} of equally old populations ($>10$ Gyr) that differ in metallicity. The difference in metallicity may indicate distinct cluster formation mechanisms for the two populations (blue or metal-poor and red or metal-rich), with small age differences allowed within the uncertainties of current age estimates. However, various GC systems associated with early-type galaxies have turned out to be complex systems showing multiple subpopulations of clusters \citep{Blom2012,Caso2013,Caso2015,Escudero2015,Sesto2016,Bassino2017,Sesto2018,Escudero2020}.

In this sense, \citet{Chies-Santos2011} have studied GC systems of elliptical (E) and lenticular (S0) galaxies using optical/NIR photometry, finding that S0 galaxies contain `blue' GCs that are most likely on average younger than those of ellipticals. This difference could be interpreted as a result of minor mergers, where part of the `blue' GCs of S0 galaxies could have formed in dwarf galaxies and later been accreted onto their current hosts. In fact, metal-poor (blue) and young GCs are preferably formed in low-mass galaxies, in the local universe \citep{Beasley2020}.
Hence, GCs associated with S0 galaxies should exhibit a range of ages and different kinematics according to their formation or accretion mechanism. 
In the Milky Way, stars of accreted dwarf galaxies present different kinematics depending on their in-fall time, where recently acquired satellites have a larger difference to the galaxy systemic velocity and extend to larger radii \citep{Rocha2012}. Similar results are found for the GC systems of NGC\,1023 and NGC\,3115 \citep{Cortesi2016,Zanatta2018}, where accreted GCs were encountered on the basis of their kinematics. In the case of NGC\,1023, the majority of rejected GCs are consistent with having blue colours.

There are very few galaxies for which a combined study of GCs (both stellar population and kinematics), stellar light and other large galactocentric tracers has been attempted, due to the cost of observing time and data collection. Nevertheless, these types of studies are of vital importance in shedding light on the formation of galaxies, since they provide information on the kinematics and stellar population of the galaxy outskirts, which are otherwise impossible to gain. 

In this work, we perform the study of the recent merger remnant E2 galaxy NGC\,4382 \citep{Kormendy2009}, located in the outskirts of the Virgo cluster, at a projected distance of 1.7 Mpc from the cD galaxy M87 \citep{Sivakoff2003,Nagino2010}.
It is relatively isolated in the outskirts of the cluster, interacting with its companion NGC\,4394. However, NGC\,4382 has also been classified as S0 due to the surrounding diffuse stellar light \citep{Gultekin2011} and as Sa because it displays bluer colours than those expected for an early-type galaxy and an inner disk with spiral patterns \citep{Sivakoff2003}. 

NGC\,4382 is also an X-ray faint early-type galaxy \citep{Fabbiano1994}. Its X-ray to optical flux ratio is among the lowest detected in E and S0 galaxies, being $\sim$100 times smaller than those of the X-ray brightest galaxies in Virgo and in the field of similar optical luminosity \citep{Kim1996}. 
According to \citet{Sivakoff2003}, 45\% of the X-ray Chandra counts within two effective radii in NGC\,4382 is attributed to diffuse gas. This high fraction of diffuse gas has also been observed in another X-ray faint S0 galaxy, NGC\,1553 \citep{Blanton2001}. \citet{Sivakoff2003} also state that the large core radius of NGC\,4382, its rounder X-ray emission than optical emission, and lower gas temperature at inner radii are consistent with a rotating interstellar gas distribution as predicted by \citet{Brighenti1996}. 

Futhermore, the galaxy presents several particular characteristics as a result of its past mergers, such as distorted and boxy isophotes \citep{Burstein1979,Ferrarese2006}, shells and ripples \citep{Schweizer1988}, excess light at intermediate radius in its brightness profile \citep{Kormendy2009}, kinematically decoupled core \citep{McDermid2004}, among others. 
Regarding its GC system, different photometric and spectroscopic studies indicate the presence of more than two subpopulations of clusters \citep{Peng2006b,Chies-Santos2011,Trancho2014,Ko2018,Ko2019,Ko2020}. In particular, \citet{Ko2018} (hereafter K18) analyzed a sample of 21 objects (20 GCs) associated with NGC\,4382 using spectra obtained by the Gemini/GMOS instrument. These authors determined that 55\% of their GCs located at $R<3$ arcmin had a mean age of $\sim$4 Gyr. Later, \citet{Ko2020} (hereafter K20), using the MMT/Hectospec camera, presented a wide field spectroscopic analysis of 89 GCs covering $R<30$ arcmin. The study of the stellar populations of the  GC system obtained in this work, unlike that estimated by K18, reveal that the different GC subpopulations have similar ages of the order of 10 Gyr, but with different metallicities. However, it is necessary to mention that the mean age and metallicity values of the GC subpopulations estimated by K20
were obtained from co-added spectra according to the colour range $(g-i)_0$ considered for each group of clusters. This choice was dictated by the low signal-to-noise ratio (S/N) of those data, which impossibilities the fit of a single spectrum of a GC. Furthermore, this approach leads to mixing of stellar populations, which can completely muddle conclusions regarding the formation history of the system.

In order to analyze the different characteristics of the GC subpopulations associated with NGC\,4382, we carried out a photometric and spectroscopic study in the inner region of the galaxy ($R<5$ arcmin), obtaining high S/N spectra of 25 GCs,  complementing with the 21 objects of K18, thus doubling the  sample of GCs with S/N $\ge$ 10 in the mentioned region.

The work is organized as follows. In Section \ref{observations}, we present the photometric and spectroscopic data and describe the reduction process. In Section \ref{phot_analysis}, we focus on the photometric analysis of the GC system and its different subpopulations. In Section \ref{spectr_analysis}, we obtained the radial velocities of the objects in our spectroscopic sample, we derive and analyze the stellar population parameters of the GCs using the full spectral fitting technique. Also, we analyze the kinematics of the system as well as the different groups. On the other hand, in this same section, we study the stellar populations of NGC\,4382 from long-slit data. Finally, in Section \ref{conclusions} we present our summary and conclusions. In this work, according to the distance modulus adopted for NGC\,4382 (see Table \ref{NGC4382}), 1 arcsec corresponds to 0.087 kpc. 

\begin{table}
    \centering
    \caption{Basic information of NGC\,4382}
    \begin{tabular}{lcl}
    \hline
        $(m-M)$        &  31.262$\pm$0.067  & \citet{Blakeslee2009}\\
        Distance (Mpc) &  17.9$\pm$0.5 & \citet{Blakeslee2009} \\
        Morph. Type & E2 & \citet{Kormendy2009}\\
        $M_V$ (mag) & -22.54 & \citet{Gultekin2011}\\
        $M_B$ (mag) & -21.34 & \citet{devaucouleurs1991}\\
        $V_\mathrm{sys}$ (km s$^{-1}$) & $729\pm2$ & \citet{Smith2000}\\
        $\sigma$ (km s$^{-1}$) & 179 & \citet{McDermid2006}\\
        $[\alpha/\mathrm{Fe}]$ & 0.12$\pm$0.06 & \citet{McDermid2006}\\
        ISM T$_X$ (keV) & 0.3 & \citet{Nagino2010}\\
        M$_{BH}$ (10$^7$ M$_\odot$) & 1.3 & \citet{Gultekin2011}\\
    \hline
    \end{tabular}
    \label{NGC4382}
\end{table}


\section{Observations and Data Reduction}
\label{observations}
\subsection{Photometric Data}
\label{photo_data}

\begin{figure}
    \centering
	\includegraphics[width=0.9\columnwidth]{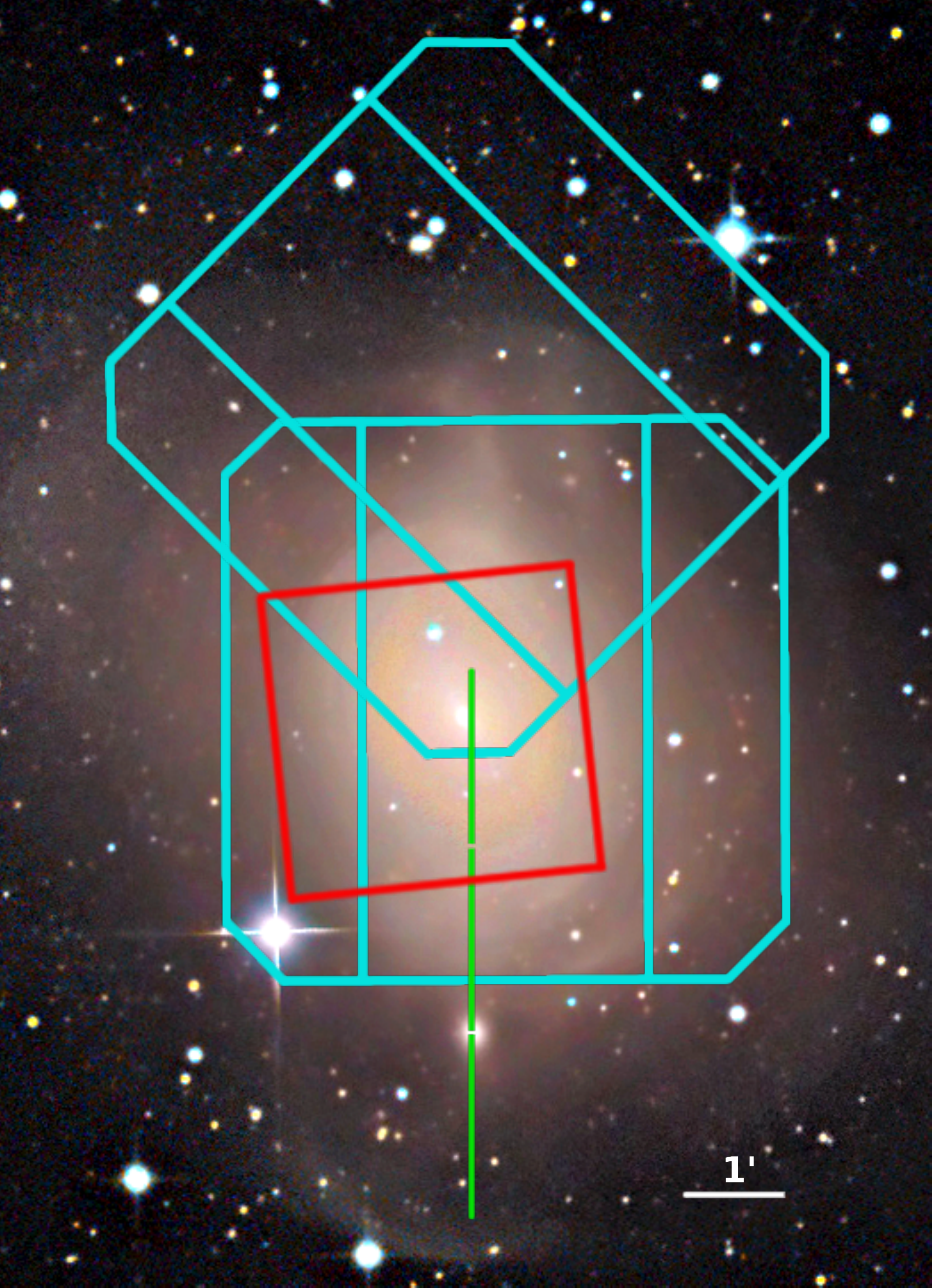}
    \caption{Colour image of NGC\,4382 taken by the Irida Observatory (https://www.irida-observatory.org/CCD/NGC4382/NGC4382.html). In cyan, the GMOS fields corresponding to the Gemini programmes GN-2014A-Q-35 (upper field) and GN-2006A-Q-81, GN-2015A-Q-207 (lower field centred on the galaxy) are shown. The red square indicates the position of the HST/ACS field used in this work. Finally, the green line indicates the position of the GMOS long-slit spectroscopic data associated with the programme GN-2009A-Q-102 (see Section \ref{spect_data}). North up and east to the left.}
    \label{fig:NGC4382_fields}
\end{figure}

Photometric data used in this work consists of images obtained from the Gemini Multi-Object Spectrograph (GMOS) camera \citep{Hook2004} mounted on the Gemini North telescope. The observed images of NGC\,4382 were taken in the $g',r',i'$ filters with a 2$\times$2 binning and correspond to the Gemini programmes GN-2014A-Q-35 (PI: A. Cortesi), GN-2006A-Q-81 (PI: O. Nakamura) and GN-2015A-Q-207 (PI: Myung Gyoon Lee). The programme GN-2014A-Q-35 comprises a dataset observed in the $g',i'$ filters located towards the north of the galaxy. On the other hand, the observations corresponding to the programmes GN-2006A-Q-81 and GN-2015A-Q-207 are centered on the galaxy, and were observed in the filters $g',i'$ and $r'$, respectively. Figure \ref{fig:NGC4382_fields} shows the orientation and distribution of the aforementioned GMOS fields. In addition, we use additional data from a GMOS field (GN-2019A-Q-903; PI: A. Zitrin) located a few degrees away from the galaxy in order to use it as a comparison field during photometric analysis (see Section \ref{sample}). Table \ref{Tab_datos} provides a summary of the observations used in this work. 

The data reduction process was carried out using Gemini/GMOS tasks within {\sc{iraf}}\footnote{IRAF is distributed by the National Optical Astronomical Observatories, which are operated by the Association of Universities for Research in Astronomy, Inc., under cooperative agreement with the National Science Foundation} (e.g. {\sc{gprepare, gbias, giflat, gireduce, gmosaic}}). To do this, we downloaded bias and flat-field images as part of the standard GMOS baseline calibrations from the Gemini Observatory Archive (GOA), in order to correct the raw science data from each programme. Subsequently, the reduced images of each field and filter were combined using the {\sc{iraf}} task {\sc{imcoadd}}. The final FWHM values are listed in Table \ref{Tab_datos}.

Finally, we complement the GMOS data with the photometric catalogue of GC candidates of NGC\,4382 in the filters $g$ (F475W) and $z$ (F850LP) of \citet{Jordan2009}. This dataset corresponds to the HST/ACS camera obtained from the ACS Virgo Cluster Survey \citep{Cote2004} (programme ID: GO-9401). However, it is necessary to consider that the ACS FoV covers a smaller area than a GMOS field. Figure \ref{fig:NGC4382_fields} shows the location of this ACS field with a red square.

\begin{table*}
\centering
\scriptsize
\caption{Set of photometric and spectroscopic observations used in this work.} 
\label{Tab_datos}
\begin{tabular}{lccccl}
\hline
\hline
\multicolumn{1}{c}{\textbf{Programme ID}} &
\multicolumn{1}{c}{\textbf{Type}} &
\multicolumn{1}{c}{\textbf{PI}} &
\multicolumn{1}{c}{\textbf{RA}} &
\multicolumn{1}{c}{\textbf{DEC}} &
\multicolumn{1}{c}{\textbf{Description}} \\
\multicolumn{3}{c}{} &
\multicolumn{1}{c}{(h:m:s)} &
\multicolumn{1}{c}{(d:m:s)} &
\multicolumn{1}{c}{} \\
\hline
GN-2006A-Q-81  & Photometry   & Nakamura         &  12:25:22.5 & 18:11:30 & GMOS images: $g'$:2$\times$150 sec; $i'$:3$\times$150 sec; FWHM:0.59-0.85 arcsec  \\ 
GN-2009A-Q-102 & Spectroscopy & Aragón-Salamanca &  12:25:24.0 & 18:09:14 & GMOS long-slit; grating B1200; slit 0.5 arcsec; 4$\times$1 binning; 4$\times$900 sec \\
GN-2014A-Q-35  & Photometry   & Cortesi          &  12:25:24.1 & 18:14:44 & GMOS images: $g'$:4$\times$120 sec; $i'$:4$\times$120 sec; FWHM:0.43-0.48 arcsec  \\
GN-2015A-Q-207 & Photometry and & Myung Gyoon Lee  &  12:25:24.1 & 18:11:29 & GMOS images: $r'$:5$\times$60 sec; FWHM:0.75 arcsec                               \\
               & Spectroscopy &                  &             &          & GMOS MOS: grating B600; slits 1 arcsec; 2$\times$2 binning; 8$\times$1800 sec     \\
GN-2016A-Q-62  & Spectroscopy & Cortesi          &  12:25:24.1 & 18:14:44 & GMOS MOS: grating B600; slits 1 arcsec; 2$\times$2 binning; 13$\times$1850 sec    \\
GN-2019A-Q-903 & Photometry   & Zitrin           &  12:12:18.5 & 27:32:55 & GMOS images; $g'$:8$\times$600 sec; $r'$:7$\times$600 sec; $i'$:8$\times$300; FWHM:0.50-0.79 \\
GO-9401           & Photometry   & C{\^o}t{\'e}     &  12:25:25.5 & 18:11:23 & ACS images: $g$:2$\times$375 sec; $z$:2$\times$560 sec                            \\ 
\hline
\end{tabular}
\end{table*}

\subsection{Spectroscopic Data}
\label{spect_data}

The spectroscopic observations in this study were performed with the GMOS-North instrument (GMOS-N) in MOS mode as part of our Gemini programme GN-2016A-Q-62 (PI: A. Cortesi). Observations were made with a B600\_G5307 grating, a 2$\times$2 binning and slit widths of 1 arcsec, with variable slit length in order to have the largest number of objects in the mask. In this case, we placed 28 slits in the mask for GC candidates of NGC\,4382. The selection of the objects was performed using the image data from the programme GN-2014A-Q-35 ($g',i'$ filters). The spectroscopic configuration produces a dispersion of 0.92 \AA\, pixel$^{-1}$, a spectral resolution of 4.7 \AA, and covers the wavelength range of $\sim$ 3900-6800 \AA, according to the position of the target in the mask. The set of exposures is made up of thirteen observations of 1850\,s at central wavelength of 500, 505 and 510\,nm, with a total time of 6.68 hours on-source. 
The reduction of the GMOS spectroscopy was carried out with the GEMINI {\sc{iraf}} package (version V2.16). We follow the same guidelines described in \citet{Norris2015} and \citet{Sesto2018}. We use the {\sc{gbias}} and {\sc{gsflat}} tasks to obtain the master bias and master flat calibration images. Subsequently, using the {\sc{gsreduce}} task, we corrected the science data by bias, overscan subtraction and flat-field correction. Wavelength calibration was performed on the copper-argon (CuAr) arc spectra using the {\sc{gswavelength}} task. This solution was applied to the science data using the {\sc{gstransform}} task. Then, the individual spectra were sky-subtracted and extracted using the {\sc{apall}} task in the {\sc{apextract}} package, and before combining they were corrected to the heliocentric system using the {\sc{dopcor}} task. The 13 spectra of each object were combined using the task {\sc{scombine}}. To flux-calibrate our data, the flux standard star Hiltner\,600 was observed. The reduction of this dataset was carried out with the same procedures mentioned above. We obtained the sensitivity function using the {\sc{gsstandard}} task in order to calibrate the standard star spectrum and the science spectra. This last step was carried out using the {\sc{calibrate}} task of {\sc{iraf}}.

\begin{figure}
	\includegraphics[width=0.99\columnwidth]{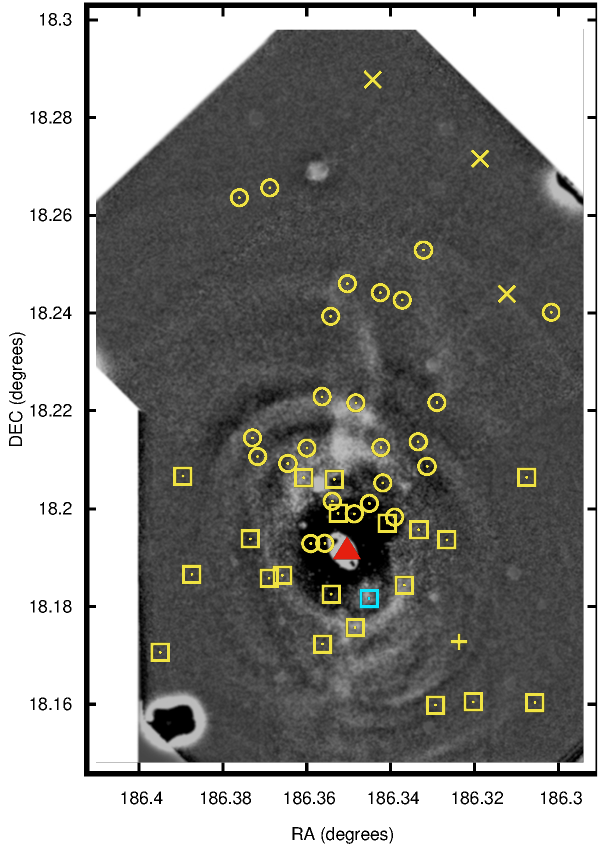}
    \caption{NGC\,4382 composite image of two GMOS fields in filter $i'$ (see Section \ref{photo_data}). The foreground and background objects were removed and the unsharp masking technique was applied to highlight the fine-structure features of the galaxy. The red triangle indicates the galactic centre. The circles and squares point to the location of GCs identify by programmes GN-2016A-Q-62 and GN-2015A-Q-207, respectively. The crosses and the plus symbol indicate the objects that were later confirmed as contaminants. The cyan square indicates the location of hypercompact cluster M85-HCC1 (see Section \ref{populations}). 
    North up and east to the left.}
    \label{fig:spatial_spectra}
\end{figure}

With the aim to increase the spectroscopic sample of objects, we complemented our spectra with those of K18
(programme GN-2015A-Q-207). This dataset, without objects in common with our sample, comprises 8 exposures of 1800\,s (4 hours on-source) and was observed with the same instrument (GMOS-N), grating (B600\_G5307), and slits (1 arcsec wide) as our Gemini programme. A complete description of them can be found in the work mentioned above. To obtain a homogeneous sample between both datasets, we downloaded the raw data corresponding to this program and we carry out the reducing process according to the aforementioned guidelines. In this way, the total sample of objects with spectra used in this work comprises 53 sources (28 objects in our mask and 25 objects corresponding to K18).
Figure \ref{fig:spatial_spectra} shows with circles the spatial distribution of the objects in our mask, and with squares the objects analyzed by K18.

Finally, we also consider the GMOS long-slit data for the galaxy corresponding to the Gemini programme GN-2009A-Q-102 (PI: Arag\'on-Salamanca). This dataset, oriented in the north-south direction of NGC\,4382 (see Figure \ref{fig:NGC4382_fields}), comprises 4 exposures of 900\,s obtained with the B1200 grating in combination with a 0.5 arcsec slit. This configuration provided a spectral resolution of $\sim$1.13\,\AA\, in the wavelength range $\sim$4300-5500\,\AA. In addition, the images were binned by 4 in the spatial direction giving a scale of 0.29 arcsec pixel$^{-1}$. Together with the observation programme, the standard star Feige\,66 was observed in order to flux-calibrate the data. A detailed description of these data can be found in \citet{Johnston2014}. The data reduction process was carried out with the same {\sc{iraf}} tasks applied for the aforementioned MOS data. 

The central position of the long-slit was placed offset from the centre of the galaxy in order to maximize the spatial coverage, as shown in Figure \ref{fig:NGC4382_fields}. Since an accurate sky subtraction is required to estimate reliable kinematic and stellar population parameters at large galactocentric radii,  we use the region of the slit furthest from the galactic centre to estimate it. The mean sky level obtained in the last 40 arcsec of the slit was then subtracted from the whole image. Finally, the four 2D reduced images of the galaxy were then combined using the {\sc{iraf}} task {\sc{lscombine}}.

\section{Photometric Analysis}
\label{phot_analysis}
\subsection{Photometry}
\label{photometry}
The detection, classification and photometry of GC candidates of NGC\,4382 was performed following the guidelines of \citet{Escudero2018,Escudero2020}. In summary, we use the combination of the {\sc{daophot}} package \citep{Stetson1987} of {\sc{iraf}} to obtain the point spread function (PSF) photometry of the detected objects, together with the versatility of SE{\sc{xtractor}} software \citep{Bertin1996} to classify. We use the parameter {\sc{class star}} of SE{\sc{xtractor}} to separate the resolved from the unresolved objects considering the limit value 0.5. The selection of this value was considered due to the distance at which the galaxy is located (see Table \ref{NGC4382}) since its GC candidates are expected to be observed as unresolved objects. Subsequently, we calibrate our photometry to the standard system using standard star fields observed on the same nights as the science data. Finally, we apply the galactic extinction coefficients of \citet{Schlafly2011} to the catalogue.

\begin{figure}
	\includegraphics[width=0.99\columnwidth]{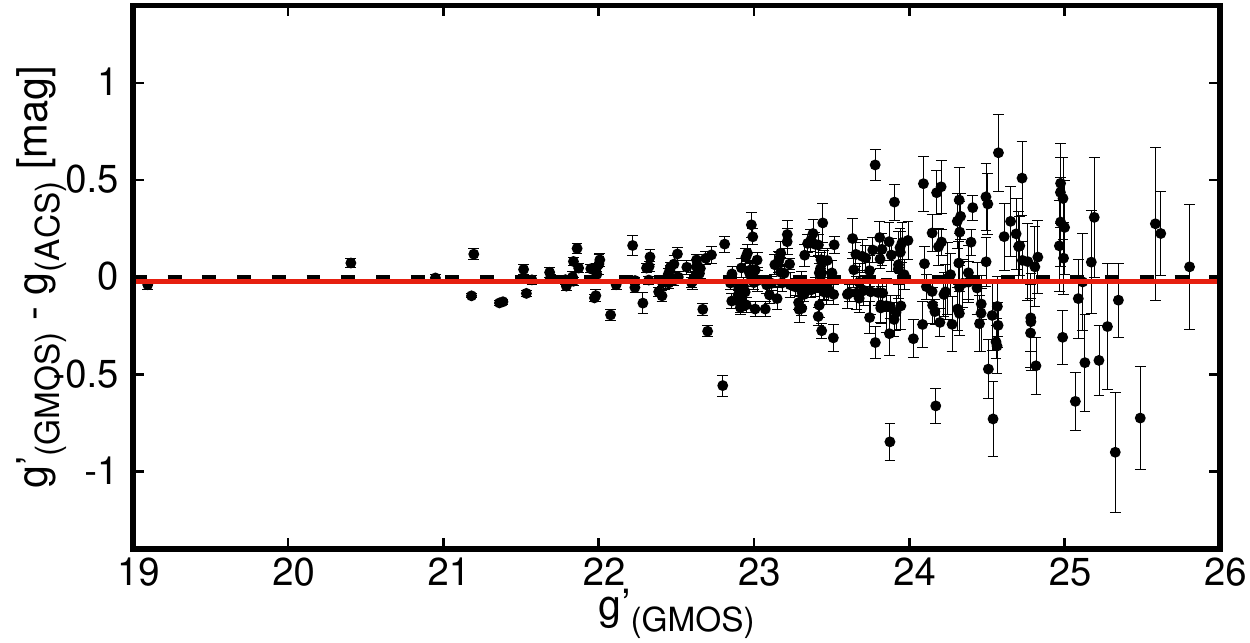}
    \caption{Comparison of magnitudes in the $g$-band between GMOS and ACS data. The dashed black line and the solid red line indicate the zero value and the difference obtained between both photometries ($\Updelta g=0.020\pm0.007$ mag), respectively.}
    \label{fig:g_comparison}
\end{figure}

Furthermore, as mentioned in Section \ref{photo_data}, we complemented the GMOS data with the ACS photometry of the GC candidates from \citet{Jordan2009}. Due to the difference between the FoVs of both cameras, only one-third of the objects in the GMOS catalogue will have information in the $g$ and $z$ filters. Figure \ref{fig:g_comparison} shows the difference in magnitudes in the $g$-band of common objects between the GMOS and ACS data. There, the excellent agreement obtained for 270 objects in both photometries is observed. The difference between them results in $0.020\pm0.007$ mag. Throughout the rest of this paper for the different analyses carried out in this work, we will use the $g'$-magnitudes of GMOS since we have a greater number of measured objects.

\subsection{GC Selection}
\label{sample}
When observing the colour-colour and colour-magnitude diagrams of GC candidates associated with early-type galaxies \citep[e.g.,][]{Harris2009,Faifer2011,Sesto2016,Escudero2017,Escudero2018}, it is observed that they are generally located around specific colours. In order to obtain an initial sample of GC candidates clean enough from contaminating objects, we considered wide colour ranges according to the filters $g',r',i'$: 
\begin{itemize}
\item $0.0<(g'-r')_0<1.05$; 
\item $0.5<(g'-i')_0<1.6$; 
\item $0.1<(r'-i')_0<0.9$ mag. 
\end{itemize}
These colour ranges were considered because NGC\,4382 is not a typical early-type galaxy, since it would present a significant number of young GCs (K18) as well as an excess of diffuse stellar clusters \citep{Peng2006b,Liu2016}. In particular, this last type of object is characterized by presenting redder colours ($1.1<(g-z)<1.6$ mag) than the typical old metal-rich GCs, as well as low luminosities ($M_V>-8$ mag) and broad half-light radii ($3-30$ pc). 
In this work, we do not consider the $z$ filter for the colour cuts since the objects with magnitudes in this filter are limited to a much smaller region than the GMOS FoV, and furthermore, these objects would already be strong GC candidates according to \citet{Jordan2009}. 

\begin{figure}
	\includegraphics[width=0.99\columnwidth]{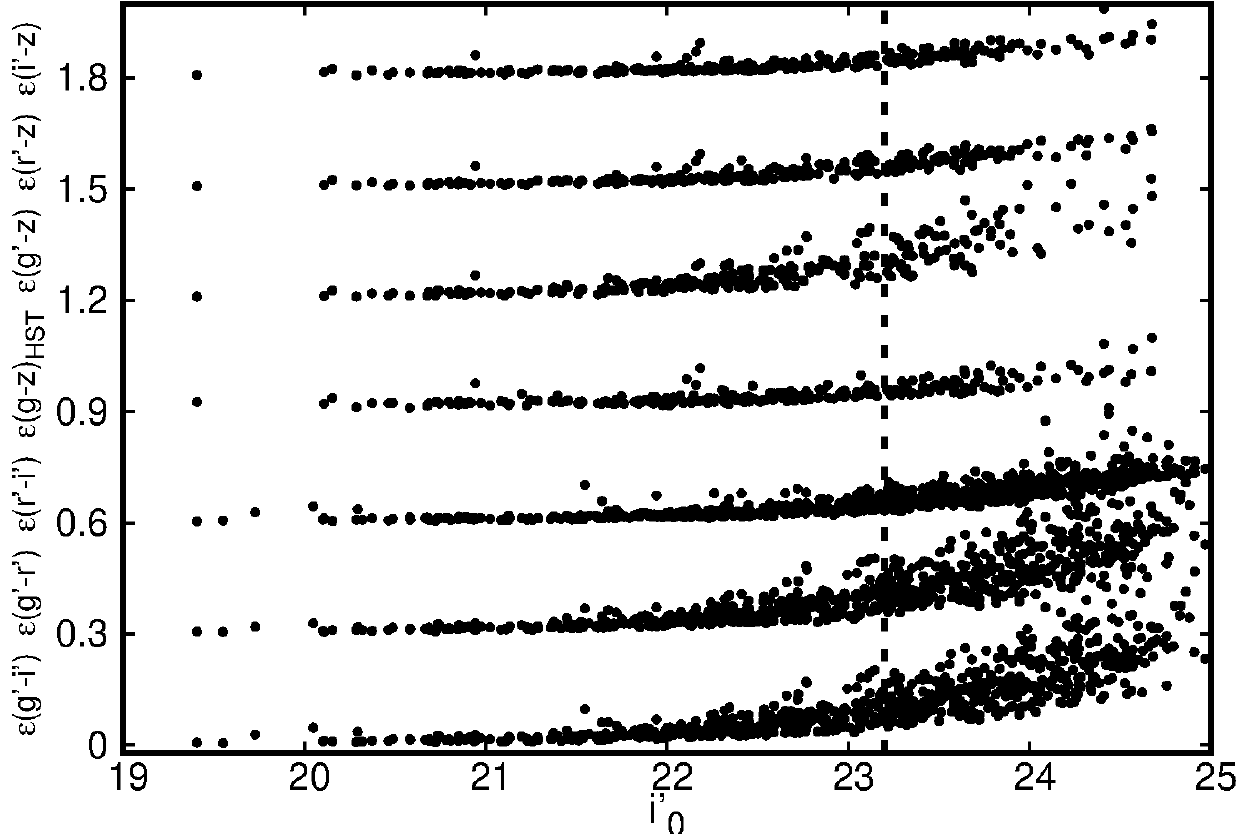}
    \caption{Photometric errors of colour indices (shifted vertically $+0.3$ mag to avoid overlapping). Vertical dashed line indicates the limiting magnitude $i'_0=23.2$ mag considered in this work.}
    \label{fig:phot_errors}
\end{figure}

\begin{figure}
	\includegraphics[width=0.99\columnwidth]{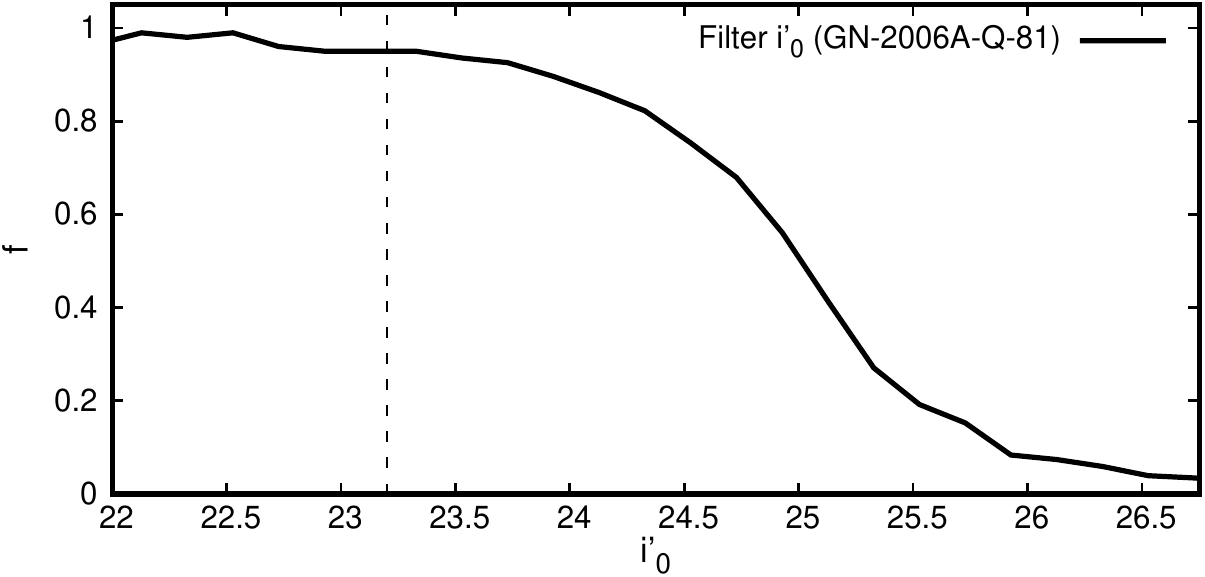}
    \caption{Completeness fraction as a function of $i'_0$ magnitude for the programme GN-2006A-Q-81 (solid line). The dashed vertical line indicates the limiting magnitude $i'_0=23.2$ mag adopted in this work (colour errors $\epsilon \sim 0.12$ mag).}
    \label{fig:completeness}
\end{figure}

The photometric quality of this initial selection of candidates is reflected in Figure \ref{fig:phot_errors}. As can be seen in the figure, those colours that involve the filter $g'$ present mean errors of the order of $\sim$0.2 at $i'_0=24$ mag. This dispersion is due to the fact that part of the objects detected in this filter corresponds to the programme GN-2006A-Q-81 in which we have just two exposures obtained under worse seeing condition than the programme GN-2014A-Q-35 (Table \ref{Tab_datos}). 
Therefore, in order to homogenize our catalogue and obtain a sample of objects with low colour errors, we consider the magnitude cutoff at $i'_0=23.2$ mag (colour errors $\epsilon \sim 0.12$ mag). This value in magnitude results in a completeness level in our sample greater than 90 per cent after performing several completeness experiments (Figure \ref{fig:completeness}). These experiments are carried out by adding 200 artificial point sources into intervals of 0.2 mag in the range $19<i'_0<27$ mag, which are then recovered using the same guidelines mentioned in Section \ref{photometry}. 

\begin{figure}
	\includegraphics[width=0.99\columnwidth]{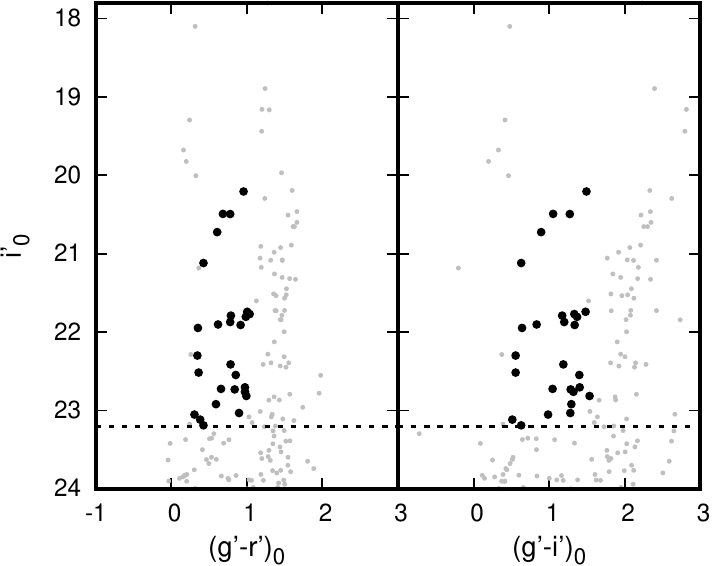}
    \caption{Colour-magnitude diagrams of the unresolved sources detected in the comparison field (grey dots). The black circles indicate the objects that present colours and magnitude within the ranges adopted for the GC candidates of NGC\,4382.}
    \label{fig:col_mag_comp}
\end{figure}

\begin{figure}
	\includegraphics[width=0.99\columnwidth]{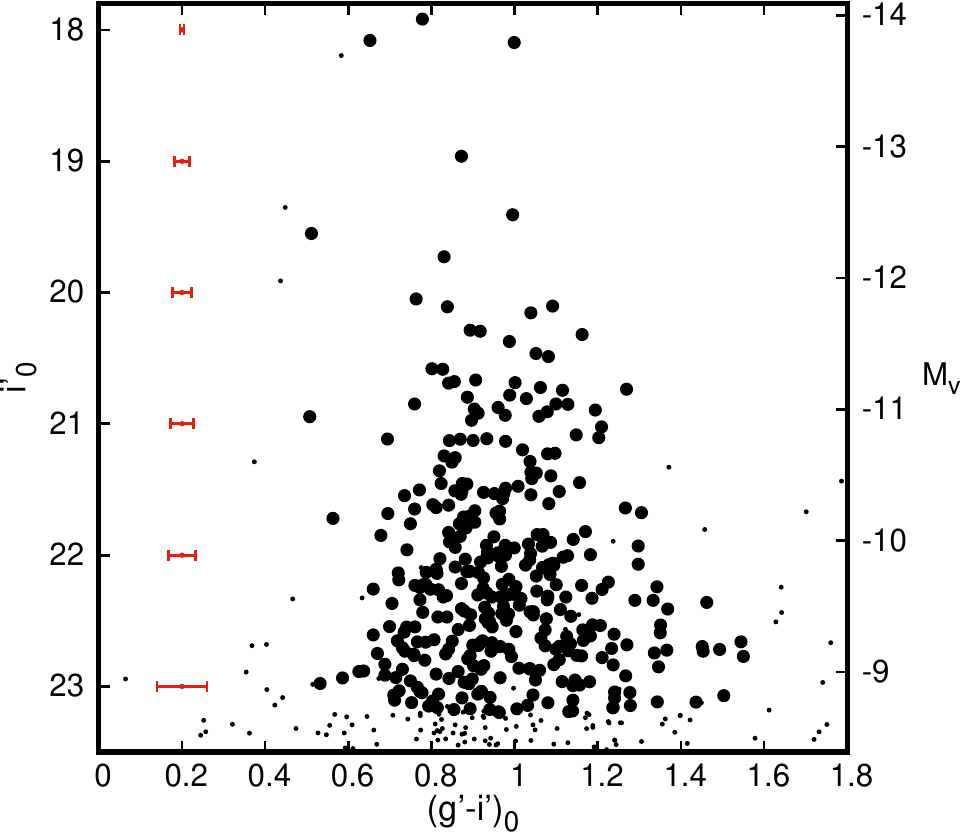}
    \caption{Colour-magnitude diagram of the unresolved sources detected in the GMOS fields (small black dots) and the final GC candidates of NGC\,4382 (black filled circles). Mean colour errors $(g'-i')_0$ are shown in red bars.}
    \label{fig:col_mag_4382}
\end{figure}

As mentioned in Section \ref{photo_data}, we used a comparison field in order to obtain a second cleanup of the GC sample for contaminating objects such as Milky Way (MW) stars or background galaxies. According to the Galactic latitude of this field ($b=81.3^\circ$), the foreground contamination should be similar to that of NGC\,4382 ($b=79.2^\circ$). We made the same cuts in colour and magnitude mentioned above and we correct the number of objects considering the ratio of areas between this comparison field and those of NGC\,4382. Figure \ref{fig:col_mag_comp} shows the colour-magnitude diagrams $i'_0$ versus $(g'-r')_0$ and $i'_0$ versus $(g'-i')_0$ with the unresolved sources detected in the comparison field. According to this field, the estimated contamination level for our final GC sample (414 objects in Figure \ref{fig:col_mag_4382}) is 9 per cent.

\subsection{GC Subpopulations}
\label{subpop}

As can be seen in Figure \ref{fig:col_mag_4382}, $(g'-i')_0$ colour distribution does not present a clear separation between the classic old metal-poor (blue) and old metal-rich (red) GC subpopulations, as is usually observed in other early-type galaxies \citep[e.g.,][]{Faifer2011,Escudero2018}. This feature observed in NGC\,4382 had already been reported in previous works such as \citet{Peng2006} and \citet{Ko2018,Ko2019}. 
Particularly in \citet{Ko2019} (hereafter K19), it is mentioned that the overall $(g-i)_0$ colour distribution of the GC system is bimodal. However, the analysis of the system in different radial bins carried out by these authors would indicate that the presence of GC candidates with intermediate colours blurs this optical bimodality towards the central region of the galaxy (R$<$4 arcmin; 20.9 kpc).

\begin{figure}
	\includegraphics[width=0.99\columnwidth]{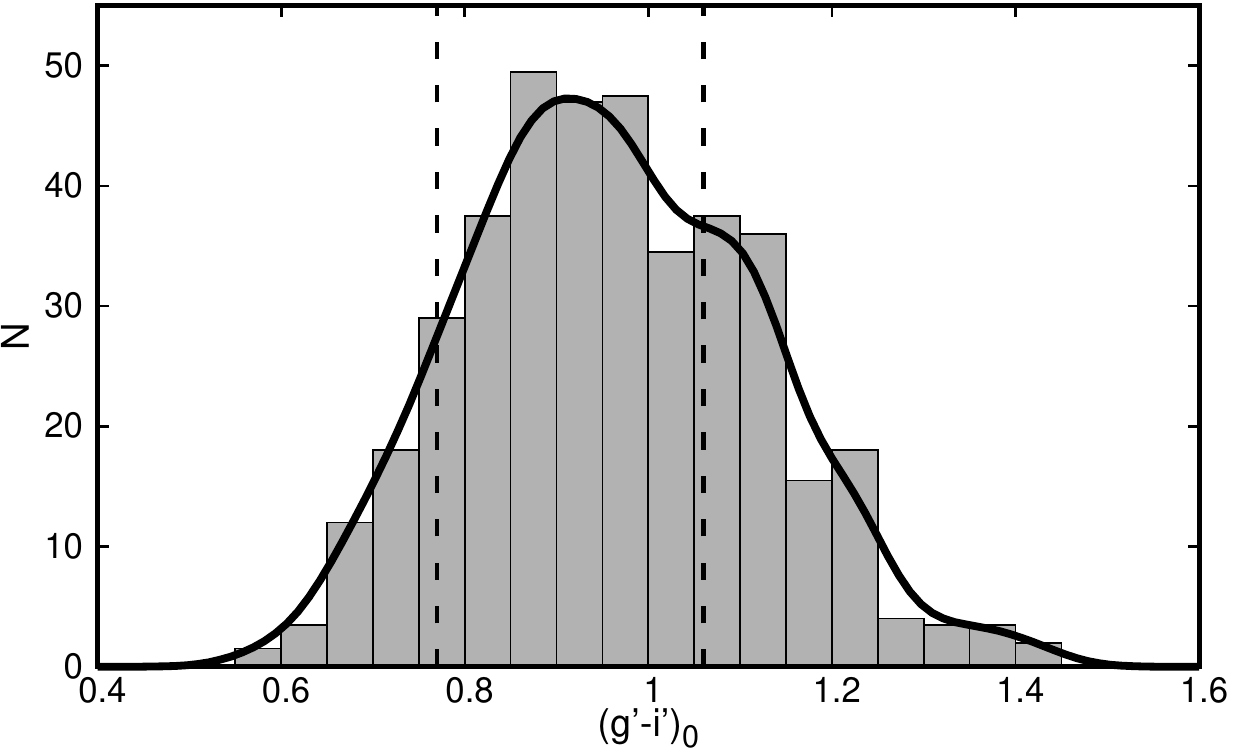}
	\includegraphics[width=0.99\columnwidth]{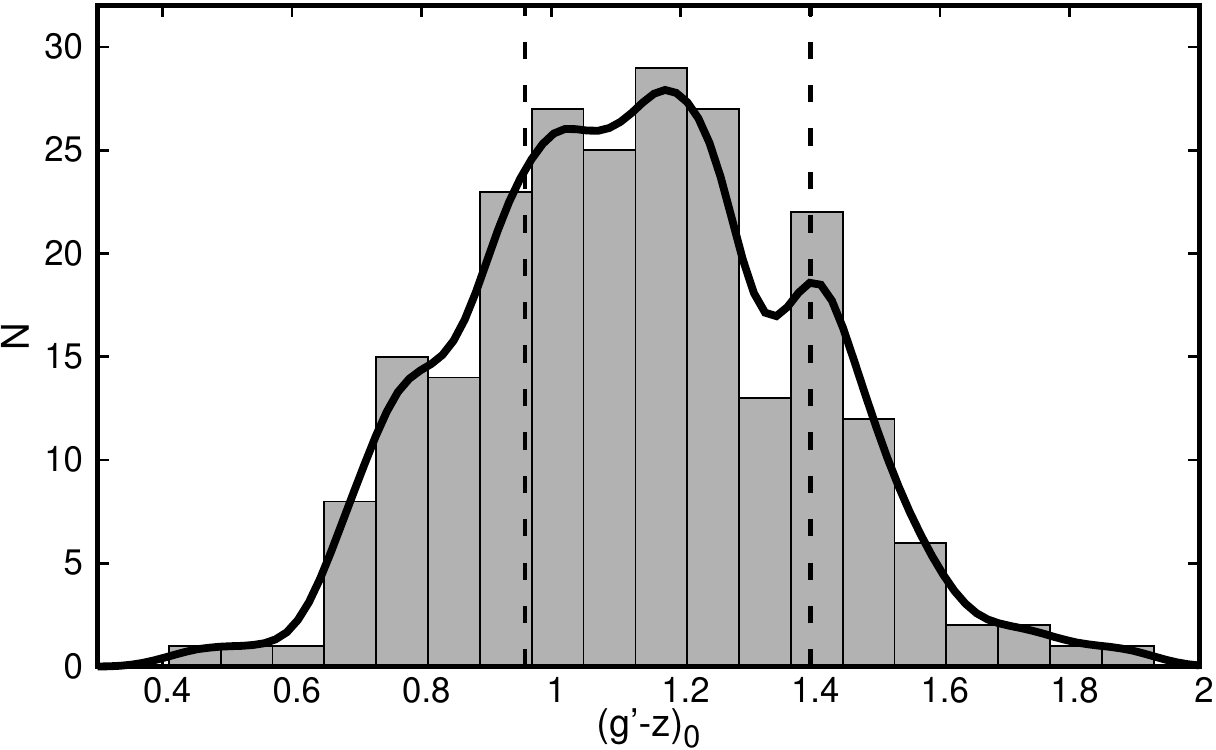}
    \caption{Background-corrected colour histograms $(g'-i')_0$ (upper panel) and $(g'-z)_0$ (bottom panel) for the GC candidates of NGC\,4382. The black line shows the smoothed colour distribution using a Gaussian kernel. The dashed vertical lines indicate the expected locations for the blue and red GC subpopulations, based on the galaxy's absolute magnitude (see text). We can already note that the colour distributions point to a more complex combination of underlying stellar populations that go beyond the bimodal distribution.}
    \label{fig:histo_gi_gz}
\end{figure}

\begin{figure*}
	\includegraphics[width=0.9\columnwidth]{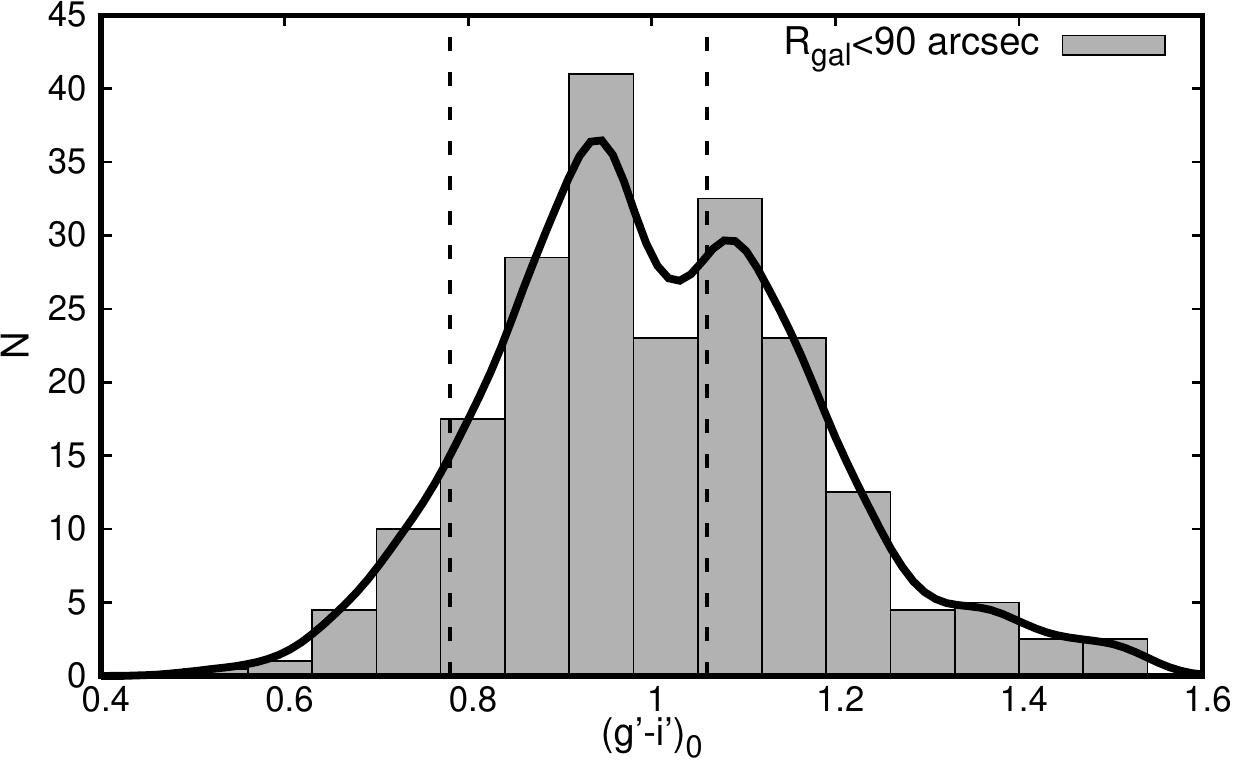}
	\includegraphics[width=0.9\columnwidth]{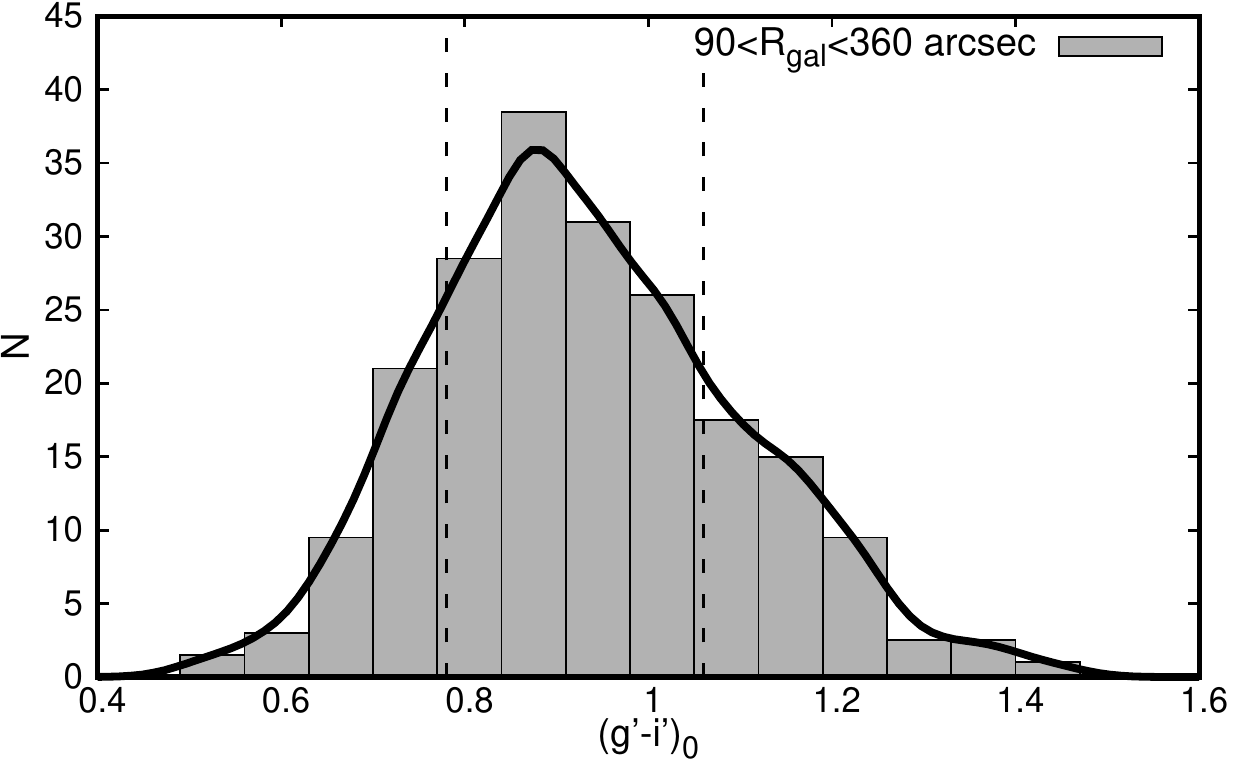}
	\includegraphics[width=0.9\columnwidth]{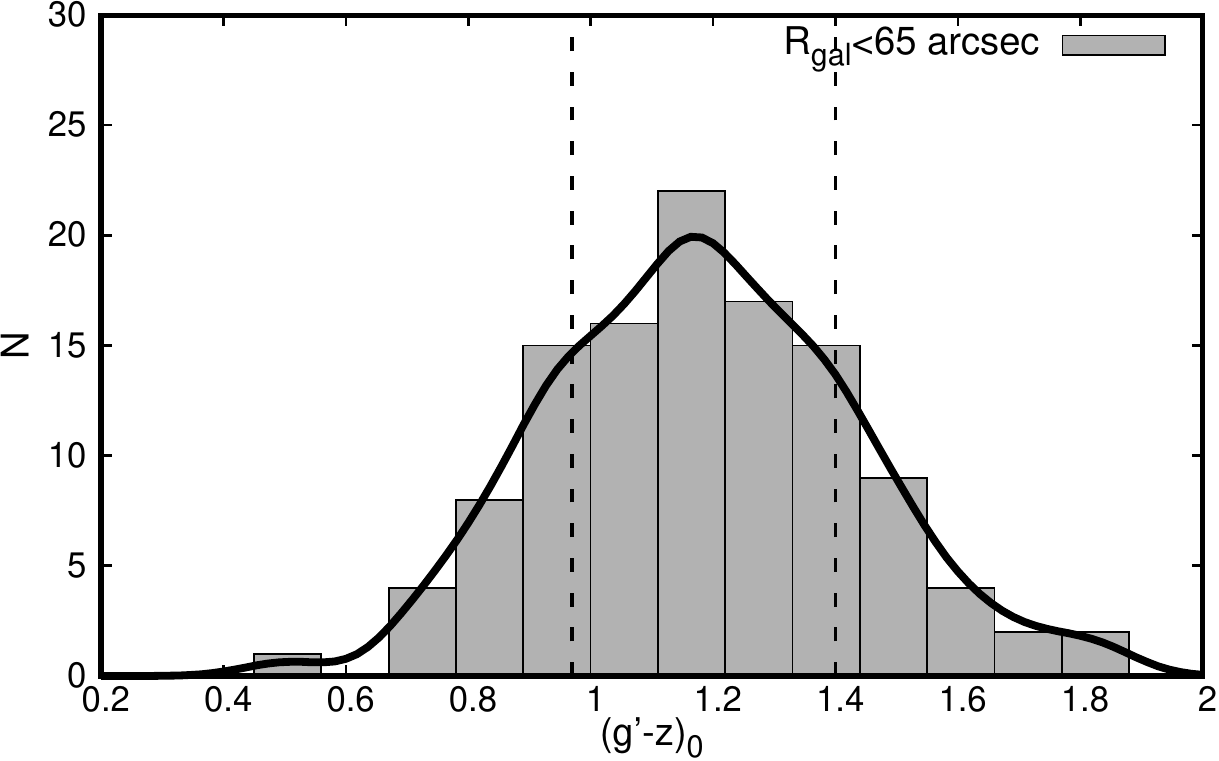}
	\includegraphics[width=0.9\columnwidth]{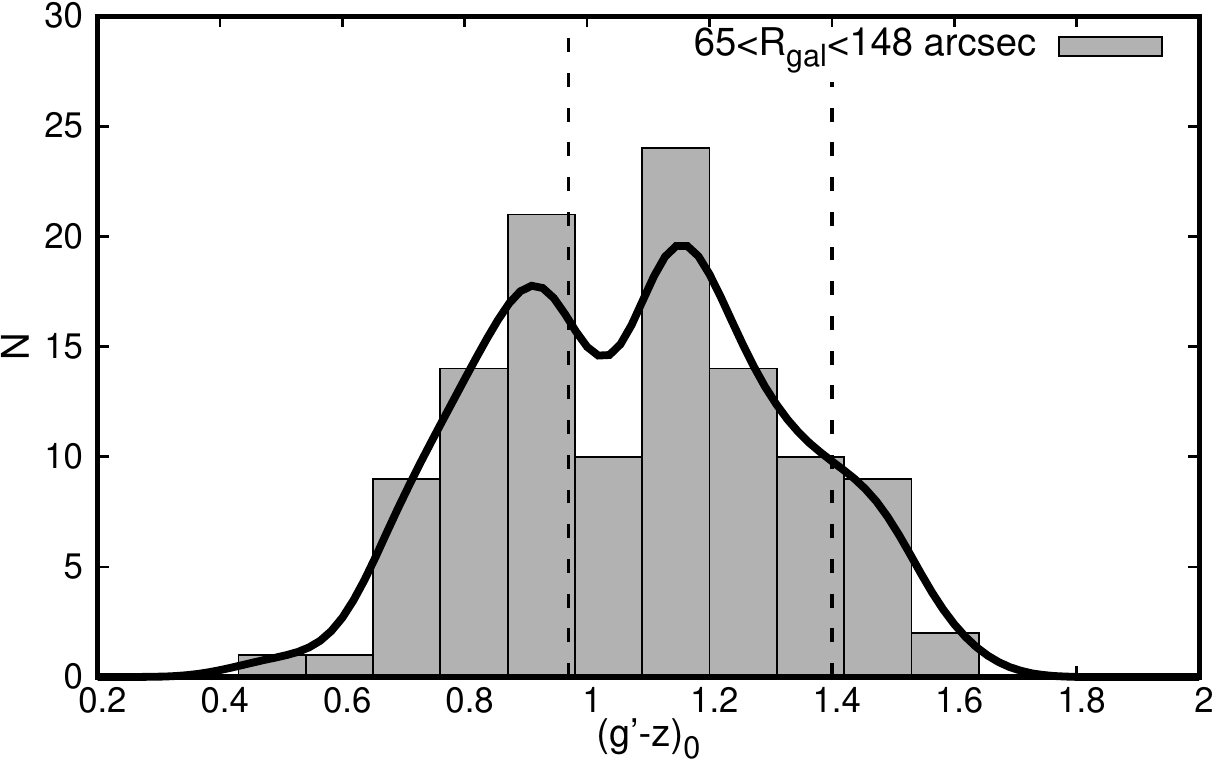}
    \caption{Background-corrected colour histograms $(g'-i')_0$ (upper panels) and $(g'-z)_0$ (bottom panels) at different galactocentric radii. Upper panels show $(g'-i')_0$ histograms at $R_\mathrm{gal}<90$ arcsec (7.8 kpc; left) and $90<R_\mathrm{gal}<360$ arcsec (right). Bottom panels show $(g'-z)_0$ histograms at $R_\mathrm{gal}<65$ arcsec (5.6 kpc; left) and $65<R_\mathrm{gal}<148$ arcsec (right). The black lines show the smoothed colour distribution using a Gaussian kernel. The dashed vertical lines indicate the expected locations for the blue and red GC subpopulations, based on the galaxy's absolute magnitude (see text).}
    \label{fig:histo_radio}
\end{figure*}

In this work, in order to analyze and separate between the different GC subpopulations of NGC\,4382, we initially construct background-corrected colour histograms $(g'-i')_0$ and $(g'-z)_0$ (Figure \ref{fig:histo_gi_gz}). To do this, we consider bin sizes of 0.05 and 0.08 mag for both histograms and Gaussian kernels with these same values for $\sigma$ to obtain the smoothed colour distributions. Using the expressions of \citet{Peng2006} and \citet{Faifer2011}, we estimate the expected mean colours of the blue and red GC subpopulations according to the absolute blue magnitude of NGC\,4382 (see Table \ref{NGC4382}), obtaining $(g'-i')_0=0.77$; $(g'-z)_0=0.97$ mag for the blue subpopulation and $(g'-i')_0=1.06$; $(g'-z)_0=1.4$ mag for the red subpopulation (vertical dashed lines in Figures \ref{fig:histo_gi_gz} and \ref{fig:histo_radio}). As seen in Figure \ref{fig:histo_gi_gz}, although the position of these peaks results in an acceptable agreement with the substructures in our histograms, mainly in the red GC subpopulation, the presence of a group of objects with intermediate colours in $(g'-i')_0\sim0.95$ and $(g'-z)_0\sim1.2$ mag stands out. Furthermore, the position of the intermediate peak in this galaxy is consistent with other GC systems that would present more than two GC subpopulations \citep{Blom2012,Escudero2020}.

Subsequently, in order to visualize and delimit the colour ranges between the different groups of clusters, we separate the catalogue into two subsamples at different galactocentric radii ($R_\mathrm{gal}$), and according to the objects with information on the colours $(g'-i')_0$ and $(g'-z)_0$. 
In the first case, considering objects with information in $(g'-i')_0$, we split the catalogue at $R_\mathrm{gal}=90$ arcsec (7.8 kpc) obtaining 222 GC candidates in both subsamples ($R_\mathrm{gal}<90$ and $90<R_\mathrm{gal}<360$ arcsec).  
On the other hand, given that we have fewer objects with information in $(g'-z)_0$, and their location is limited to the ACS field, we split the catalogue at $R_\mathrm{gal}=65$ arcsec (5.6 kpc), where we obtained 115 GC candidates for both subsamples ($R_\mathrm{gal}<65$ and $65<R_\mathrm{gal}<148$ arcsec).
It is worth mentioning that the regions corresponding to $R_\mathrm{gal}>90$ and $R_\mathrm{gal}>65$ arcsec of both colour samples have not been corrected for areal completeness.
Figure \ref{fig:histo_radio} shows the histograms for each subsample according to the colour considered.
As can be seen in the figure, the $(g'-i')_0$ histogram towards $R_\mathrm{gal}<90$ arcsec (7.8 kpc) shows the presence of the red GC subpopulation at $(g'-i')_0\sim1.1$ mag, and a clear peak with intermediate colour at $(g'-i')_0\sim0.93$ mag. Outside this region ($90<R_\mathrm{gal}<360$ arcsec), the histogram becomes unimodal with a peak at $(g'-i')_0\sim0.85$ mag, probably due to a mixture between these intermediate clusters and the classic blue subpopulation which begins to be significant towards larger galactocentric radii. In the case of $(g'-z)_0$ histograms, since the candidates are restricted to a smaller area around the galactic centre, the clear domain of objects with intermediate colours $(g'-z)_0\sim1.2$ mag is observed.

It is clear that the galaxy has more than two families of GCs in optical colours, so a correct separation between them is necessary to correctly analyze their properties. In this context, we tackle the analysis using the same procedure as in \citet{Escudero2020}, according to the filters available in our catalogue, which allows us to better disentangle the underlaying multiple subpopulations present in the global GC system of the galaxy. In this case, we use the joint information of the colour indices $(g'-i')_0$ and $(g'-r')_0$ in the colour-colour diagram (see Section \ref{sample}), together with the Gaussian Mixture Model (GMM) code through the open-source machine learning library {\sc{scikit-learn}}\footnote{http://scikit-learn.org} for python. A detailed description of the method used here can be found in \citet{Escudero2020}. The choice of the aforementioned colours reduces the total number of GC candidates, since we do not have information in the filter $r'$ of a certain region of the GMOS field associated with the programme GN-2014A-Q-35. However, the final number of objects with these colours (350 GC candidates), and uncorrected for contamination, is still significant. For this type of analysis, we do not consider the colours that involve the $z$ filter because the final sample would be considerably reduced by 40 per cent.
It is necessary to mention that K19 carried out a similar analysis using the classic GMM code \citep{Muratov2010} but using only the information in colour $(g-i)_0$. 

To estimate the optimal number of groups in our sample, we use two probabilistic statistical measures, the Akaike \citep[AIC;][]{Akaike1974} and Bayesian Information Criterion \citep[BIC;][]{Schwarz1978}. These two parameters allow us to estimate the quality of each used model, relative to each of the other models. In both parameters, the model with the lowest AIC and BIC is preferred. In this case, we consider a maximum number of 9 components/subpopulations ($K$) in the sample. 

\begin{figure}
	\includegraphics[width=0.97\columnwidth]{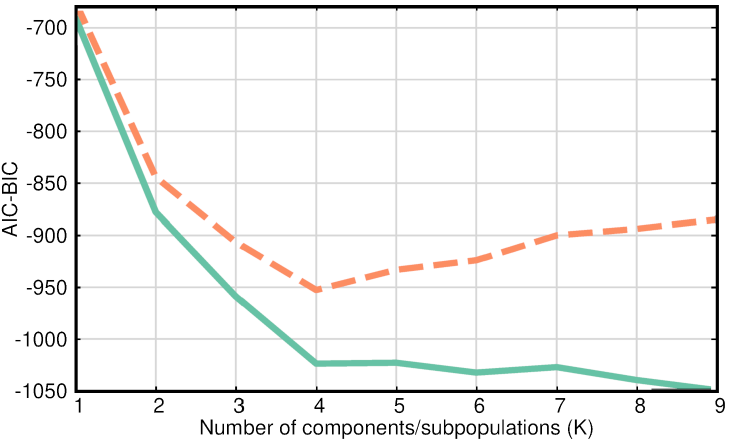}
	\includegraphics[width=0.98\columnwidth]{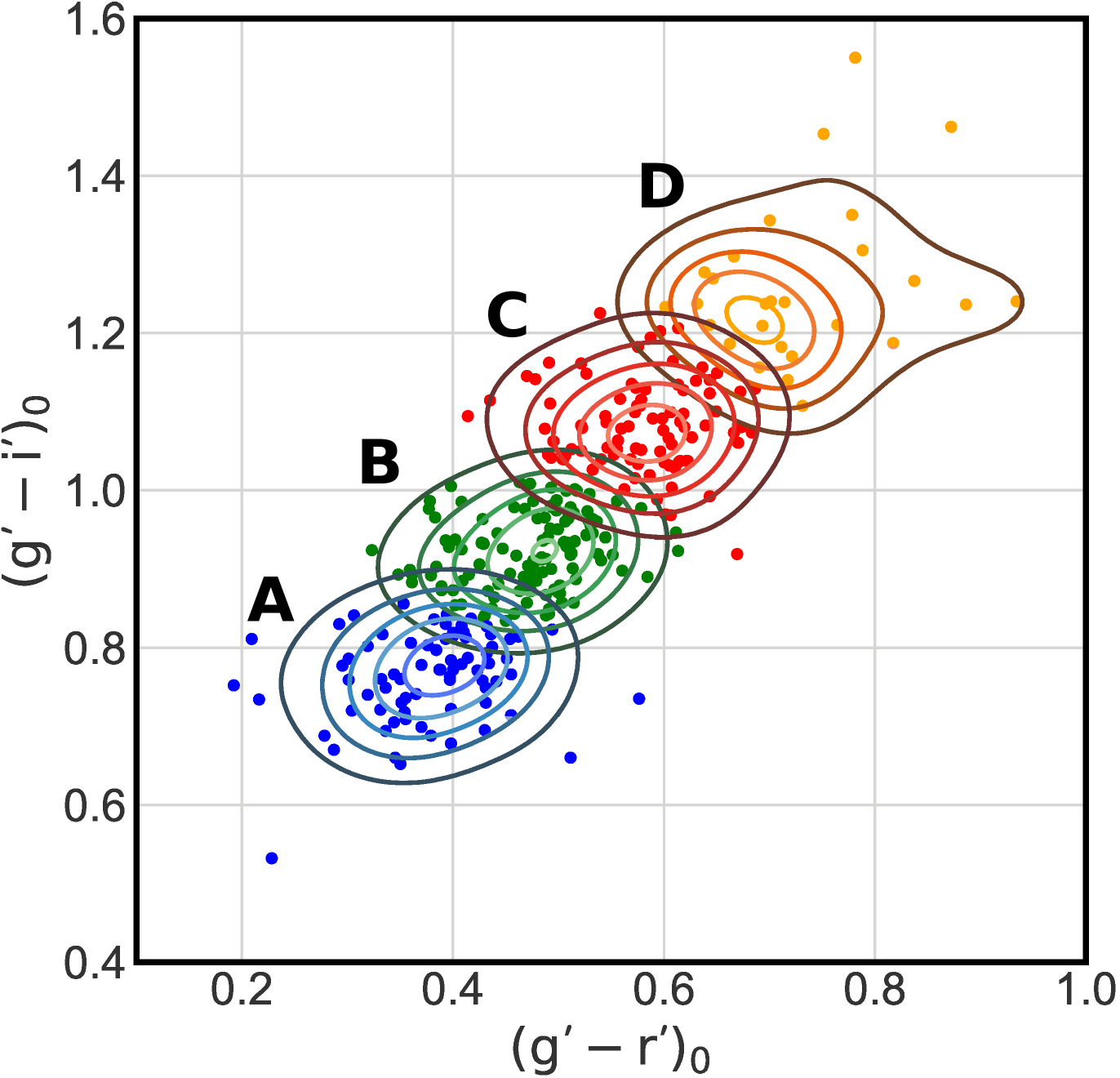}
    \caption{Upper panel: Akaike Information Criterion (AIC; solid line) and Bayesian Information Criterion (BIC; dashed line) as a function of the number of components/subpopulations ($K$). These two parameters were used to statistically identify the optimal number of distinct groups of GCs in our sample. As can be seen, both parameters point to an optimal number of 4, indicated by the break in the curves. Bottom panel: colour-colour diagram with the objects assigned to each group with different colours according to GMM. The solid lines indicate the density distribution of each group using bivariate kernel density estimate with bandwidth 0.05 mag.}
    \label{fig:AIC_BIC}
\end{figure}

We ran both statistical tests (AIC, BIC) on the sample of GC candidates clean of contaminants. This sample was generated as follows; $(g'-i')_0$ colour histograms were created for the original sample of GCs and the comparison field, correcting the latter for areal completeness. Subsequently, objects were randomly removed in each bin of the candidates' histogram according to the number of counts obtained from the comparison field. 
As observed in Figure \ref{fig:AIC_BIC} (upper panel), both statistics show a break in the curve at $K=4$, indicating that this value would be the minimum number of groups in our GC sample. 

\begin{figure*}
	\includegraphics[width=2\columnwidth]{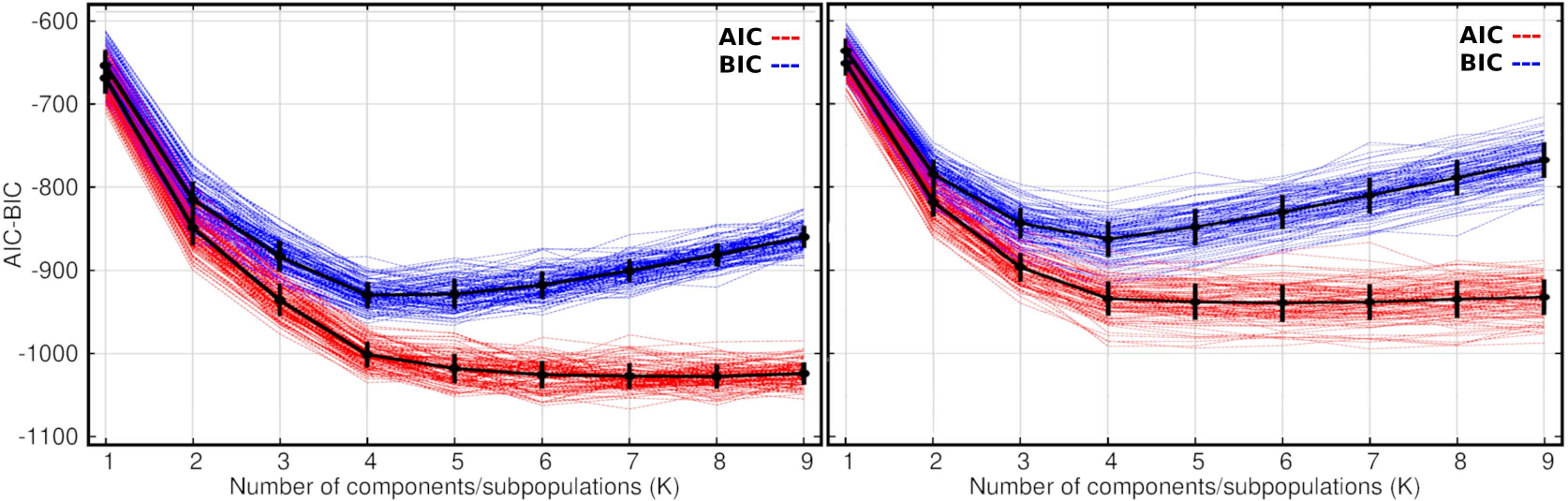}
    \caption{AIC and BIC values as a function of the number of components/subpopulations ($K$) obtained for 100 experiments considering AIC and BIC values as a function of the number of components/subpopulations ($K$) obtained on 100 experiments. The left panel shows the results obtained on 100 samples of GC candidates cleaned of contaminants as mentioned in the text. The right panel shows the results obtained considering 100 samples of GC candidates where the colours $(g'-i')_0$ and $(g'-r')_0$ of the objects within their uncertainties were randomly changed in each sample. The black lines indicate the mean and standard deviation values of the 100 experiments.}
    \label{fig:AIC_BIC_test}
\end{figure*}

However, to confirm this result, we carried out 100 experiments for each of the two statistical tests as follows. 
In each of these experiments, objects were randomly removed according to the aforementioned procedure, obtaining different samples in each run. The left panel of Figure \ref{fig:AIC_BIC_test} shows the 100 curves obtained by both tests, with the mean value indicated by a solid line. As can be seen, the mean AIC and BIC values indicate that the optimal number of groups is four. 
Subsequently, in order to verify whether the colour uncertainties could affect the results, we generate 100 new experiments randomly changing in each run the colour values $(g'-i')_0$ and $(g'-r')_0$ of the objects within their uncertainties. The right panel of Figure \ref{fig:AIC_BIC_test} shows that although the dispersion of the 100 AIC and BIC curves increases, the general trend shows that 4 groups would be optimal according to both tests.
Therefore, for the following analyzes, we consider that the sample of GC candidates of NGC\,4382 is made up of 4 different groups or subpopulations.

In order to separate these four groups in the colour plane $(g'-i')_0$ versus $(g'-r')_0$ of our GC sample, we ran the GMM code. Figure \ref{fig:AIC_BIC} (bottom panel) and Table \ref{Table_GMM} show the four groups assigned by GMM and the values obtained for each of them, respectively. Unlike the analysis carried out in the colour histograms where a clear separation between the different groups was not obtained, the analysis performed in the $(g'-i')_0$-$(g'-r')_0$ colour plane indicates colour peaks for the blue (group A) and red (group C) GC candidates similar to those expected in other early-type galaxies. 
In addition to the two usual GC subpopulations, a group with intermediate colours (group B) and a group towards redder colours (group D) are identified.  
In particular, when comparing the colour peak positions obtained in this work (Table \ref{Table_GMM}) with the values of the blue and red peaks obtained by K19 transformed to the AB system of SDSS\footnote{http://www.cadc-ccda.hia-iha.nrc-cnrc.gc.ca/en/megapipe/docs/filtold.html} ($(g-i)_0$=0.73 and $(g-i)_0$=0.99 mag, respectively), a small difference is observed in both values. In that work, the authors mention that the mean colour of the red GC subpopulation is towards bluer colours probably due to the presence of the intermediate group, which we determined in a clearer way in our colour-colour analysis.

\begin{table}
\caption{Parameters obtained by GMM. $p_{i}$: percentage of the number of GC candidates associated with each group. $w_{i}$: mixture weights. $\vec{\mu}_{i}$: central vectors of the groups. $\vec{\Sigma}_{i}$: covariance matrices.}
\label{Table_GMM}
\begin{center}
\begin{tabular}{c|ccc}\hline\hline
	Parameter  &   value   &  \\
	\hline
     $p_{A}$ & $30\pm4$  \\
     $p_{B}$ & $34\pm6$ \\
     $p_{C}$ & $28\pm7$  \\
     $p_{D}$ & $8\pm4$  \\
	\hline        
 	 $w_{A}$ & $0.26\pm0.01$ \\
     $w_{B}$ & $0.35\pm0.01$ \\
	 $w_{C}$ & $0.27\pm0.01$ \\
     $w_{D}$ & $0.12\pm0.02$ \\
	\hline
                        &  $(g'-r')$ ~~~ $(g'-i')$  &	\\
	$\vec{\mu}_{A}$	& $\left(\begin{array}{l l} 0.38\pm0.01 & 0.77\pm0.01  \end{array}\right)$	\\
	$\vec{\mu}_{B}$	& $\left(\begin{array}{l l} 0.47\pm0.01 & 0.92\pm0.01  \end{array}\right)$	\\
	$\vec{\mu}_{C}$	& $\left(\begin{array}{l l} 0.57\pm0.01 & 1.07\pm0.01  \end{array}\right)$	\\
    $\vec{\mu}_{D}$	& $\left(\begin{array}{l l} 0.69\pm0.02 & 1.21\pm0.02  \end{array}\right)$	\\
    \hline
    
	$\vec{\Sigma}_{A}$  & $\left(\begin{array}{r r} 0.005  & 0.001 \\
                                                    0.001  & 0.005 \\
                                                 \end{array} \right)$ \\
    $\vec{\Sigma}_{B}$  & $\left(\begin{array}{r r} 0.003 & 0.001 \\
                                                    0.001 & 0.003 \\
                                                 \end{array} \right)$ \\
    $\vec{\Sigma}_{C}$  & $\left(\begin{array}{r r} 0.003 & 0.001 \\
                                                    0.001 & 0.004 \\
                                                 \end{array} \right)$ \\
    $\vec{\Sigma}_{D}$ & $\left(\begin{array}{r r} 0.010 & 0.007 \\
                                                   0.007 & 0.015 \\
                                                \end{array} \right)$ \\
	\hline \hline
\end{tabular}
\end{center}
\end{table}


\subsection{Spatial Distribution}
\label{spatial}
\begin{figure}
	\includegraphics[width=0.99\columnwidth]{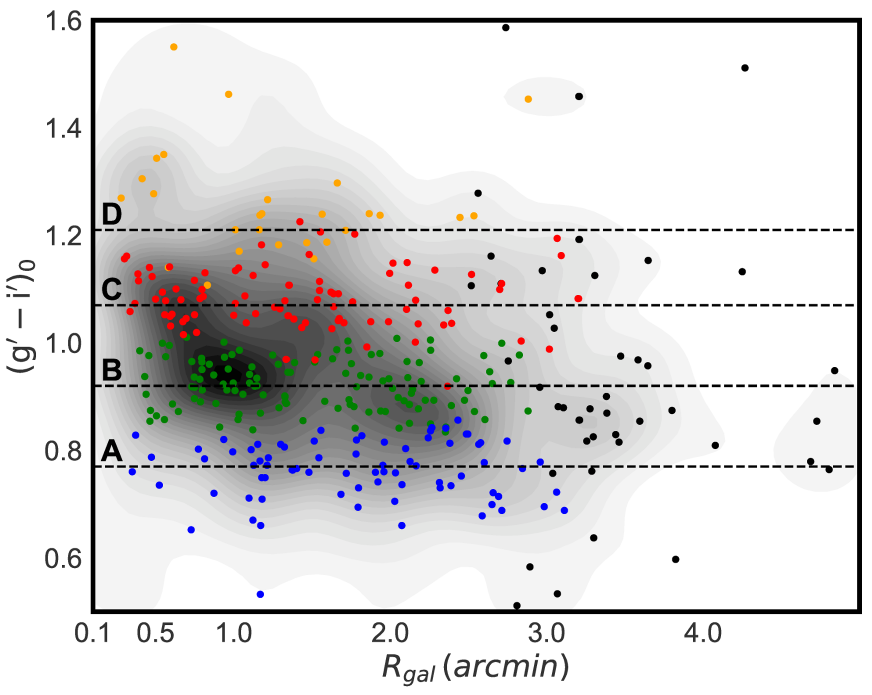}
    \caption{Grayscale representation of the smoothed colour $(g'-i')_0$ versus galactocentric radius diagram of the GC candidates of NGC\,4382. Blue, green, red and orange overplotted points indicate GC candidates assigned by GMM to groups A, B, C and D, respectively. Horizontal dashed lines indicate the mean colour $(g'-i')_0$ of each group. The black points are the GC candidates without information in the $r'$ filter.}
    \label{fig:col_dist}
\end{figure}

According to the separation performed by GMM, we plotted the $(g'-i')_0$ colour and the smoothed colour distribution of the different GC groups as a function of the projected galactocentric radius (Figure \ref{fig:col_dist}). 
In order to increase the  photometric sample towards larger $R_\mathrm{gal}$, we included in the figure the GC candidates (black dots) from the whole GMOS sample (programmes GN-2014A-Q-35 and GN-2006A-Q-81). As mentioned in Section \ref{subpop}, these objects have not been considered in the GMM analysis due to the lack of information in the $r'$ filter. 

Figure \ref{fig:col_dist} shows that the blue candidates (group A) are homogeneously distributed around the galaxy in the entire range $0<R_\mathrm{gal}<5$ arcmin ($0<R_\mathrm{gal}<26.1$ kpc). For their part, the red GC candidates (group C) show a distribution throughout the GMOS field but with a higher concentration towards $R_\mathrm{gal}<1$ arcmin ($R_\mathrm{gal}<5.2$ kpc). 
On the other hand, when observing the GC candidates with intermediate colours (group B), the presence of two clumps located at different galactocentric radii ($R_\mathrm{gal}\sim1$ arcmin (5.2 kpc) and 2.1 arcmin (10.9 kpc), respectively), stands out. Finally, although the group associated with GC candidates with the reddest colours (group D) presents a low number of objects, they seem to show a similar spatial distribution to the red ones, with a small subgroup located near the galactic centre ($R_\mathrm{gal}<0.6$ arcmin; $R_\mathrm{gal}<3.1$ kpc) while the rest are distributed  between $1<R_\mathrm{gal}<2$ arcmin ($5.2<R_\mathrm{gal}<10.4$ kpc). 

\begin{figure*}
	\includegraphics[width=1.9\columnwidth]{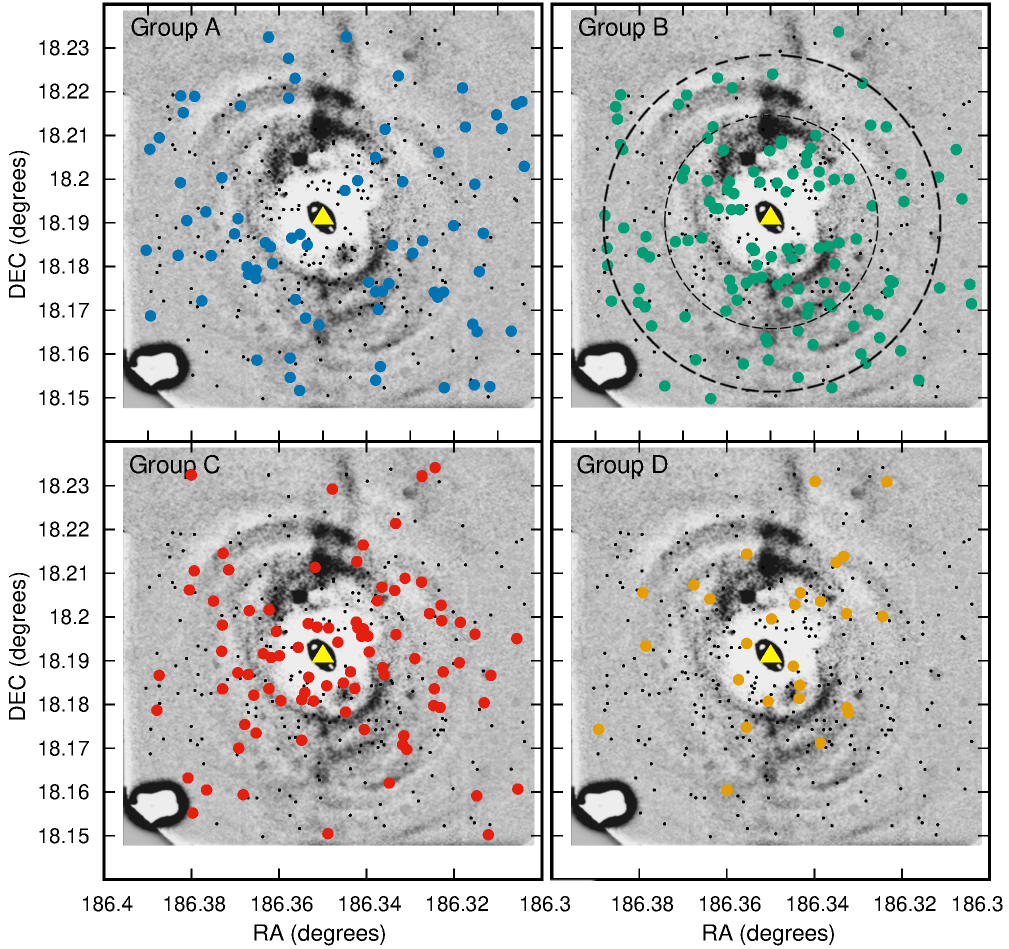}
    \caption{Spatial distribution of the 4 groups of GC candidates identified by GMM. The blue, green, red and orange points correspond to the GC subpopulations blue (upper left), intermediate (upper right), red (bottom left) and that with the reddest colours (bottom right), respectively. The black dashed lines (upper right panel) indicate the area ($1.5<R_\mathrm{gal}<2.4$ arcmin; $7.8<R_\mathrm{gal}<12.5$ kpc) where the second clump of intermediate GC candidates is located (see text). The yellow triangle indicates the centre of the galaxy. Negative unsharp-masking image in the $i'$-filter was superimposed for better visualization of the shell structures of the galaxy (dark regions) and the spatial distribution of the GC candidates. The structure located on the lower-left corner corresponds to the wavefront sensor of the GMOS instrument.}
    \label{fig:distrib_espacial}
\end{figure*}

Figure \ref{fig:distrib_espacial} shows the projected spatial distribution of the different GC groups together with the shell structures of the galaxy. Only those objects classified by GMM are shown in this figure. As mentioned above, the intermediate GC candidates (group B) show two clumps at different galactocentric radii. In particular, when observing the spatial distribution of the objects that make up the second clump ($1.5<R_\mathrm{gal}<2.4$ arcmin; $7.8<R_\mathrm{gal}<12.5$ kpc), initially a specific concentration of them would not be observed. However, some of these objects overlap in regions where different stellar shells are found. 
The black dashed circles in Figure \ref{fig:distrib_espacial} indicate the region where these objects are located. The analyses carried out by \citet{Lim2017} and \citet{Fensch2020} on the shell galaxy NGC\,474, with similar characteristics to NGC\,4382, have shown the spatial and kinematic correlation between some GCs and these stellar streams and shells. Furthermore, it is worth mentioning that the bulk of the intermediate GCs are located in the region in which an unusual "extra halo" ($0.5<R_\mathrm{gal}<3$ arcmin; $2.6<R_\mathrm{gal}<15.6$ kpc) is observed in the galaxy's brightness profile \citep{Kormendy2009}. Therefore, a spectroscopic analysis is necessary to confirm or rule out the possible association between these structures and the intermediate GC candidates of NGC\,4382. This type of analysis will be carried out in detail in a future work (Cortesi et al. in prep). 

\begin{figure}
	\includegraphics[width=0.99\columnwidth]{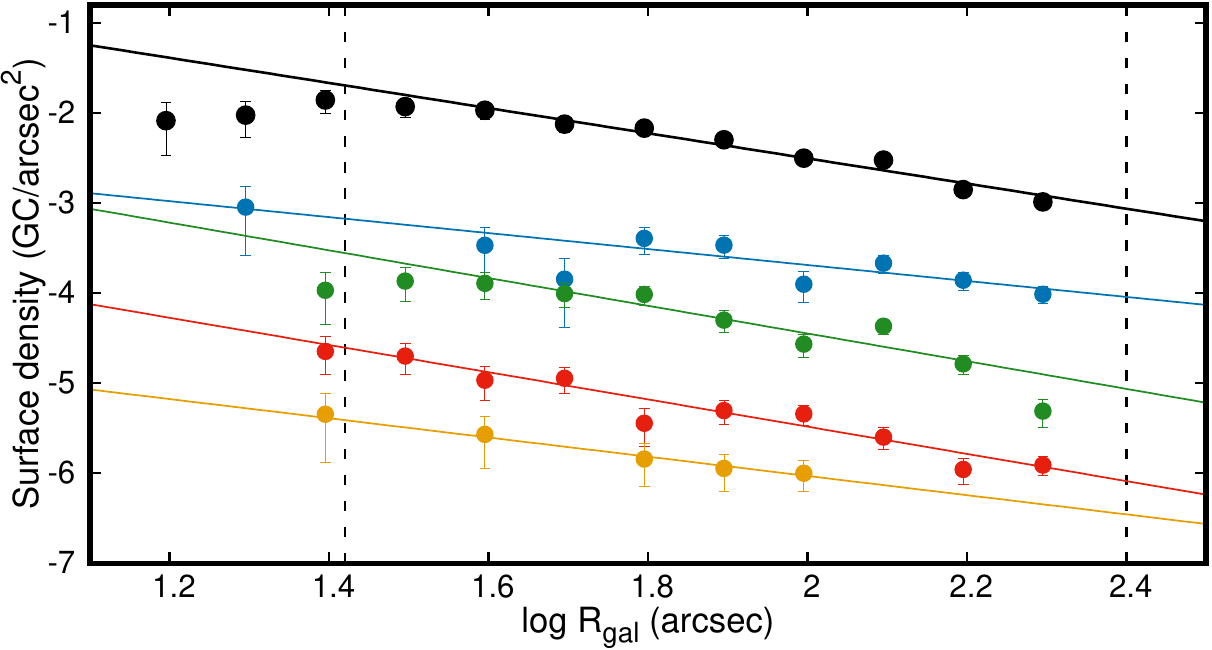}
    \caption{Projected density profiles for the GC system (black circles) of NGC\,4382, and for the blue (blue circles), intermediate (green circles), red (red circles) and redder (orange circles) subpopulations. The profiles of the subpopulations were vertically shifted to avoid overlap. The solid lines indicate the fits made on each profile considering a power-law function. Vertical dashed lines indicate the range considered for the fits.}
    \label{fig:densidad}
\end{figure}

At first glance, some of the GC subpopulations seem to show an elongated projected spatial distribution, mainly in the red (east-west direction) and intermediate (north-south direction) candidates. Therefore, in order to quantify the radial and azimuthal distribution of the four GC groups identified by GMM, we initially obtained the one-dimensional radial distribution of each of them, performing counts of objects in circular rings of step $\Delta\,log\,r=0.1$. During this process, each ring was corrected for effective area. Subsequently, we fit a power-law function ($log\,\sigma_{GC}=a+b\,log(r)$) on each profile excluding the points at $R_\mathrm{gal}<26$ arcsec ($<2.2$ kpc), in order to avoid possible incompleteness effects of the photometric sample towards the inner region of the galaxy. The obtained values are listed in Table \ref{Tabla_distr_espacial}, and Figure \ref{fig:densidad} shows the profiles with their respective fits. In agreement with that observed in Figure \ref{fig:distrib_espacial}, the intermediate (group B) and the red GC candidates (group C) show a higher concentration towards the galaxy compared to the blue subpopulation (group A). The same trend is also observed in those objects with the reddest colours (group D). In particular, the projected density profile of group B shows a wave-like behaviour, due to the presence of the two clumps of objects mentioned above, located at $R_\mathrm{gal}\sim1$ (5.2 kpc) and 2.1 arcmin (10.9 kpc).

\begin{figure}
	\includegraphics[width=0.88\columnwidth]{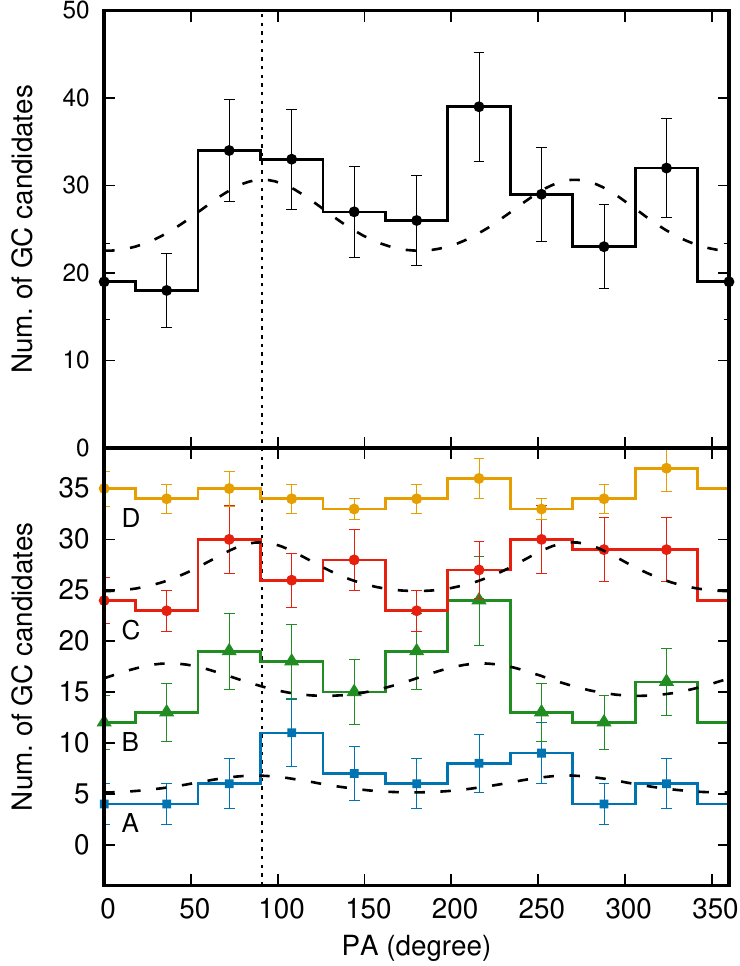}
    \caption{Azimuthal distribution of the GC system (upper panel) and of the blue, intermediate, red and redder colours subpopulations (lower panel; groups A, B, C and D, respectively). The dashed lines indicate the fits made on each sample. The histograms for groups B, C and D were shifted vertically by adding 5, 19 and 32 to the counts to avoid overlapping. The vertical dotted line indicates the $PA$ of the GC system.}
    \label{fig:acimutal}
\end{figure}

On the other hand, to analyze the azimuthal distribution, we perform object counts as a function of the position angle (measured from north to east) within the circular ring $0.25<R_\mathrm{gal}<2.5$ arcmin ($1.3<R_\mathrm{gal}<13.1$ kpc) centred in the galaxy. These values were considered to avoid correction for completeness of objects in the central region and for areal completeness of the wedges. Figure \ref{fig:acimutal} shows the histograms for the total GC sample, and for the different groups identified by GMM.
In order to determine the ellipticity ($\epsilon$) and position angle ($PA$) of each sample, we fit the \citet{McLaughlin1994} expression given by:

\begin{equation}\label{eq:acim}
  \sigma(R,\theta)=kR^{-\alpha}[cos^2(\theta-PA)+(1-\epsilon)^{-2}sin^2(\theta-PA)]^{-\alpha/2},
\end{equation}

\noindent where $\sigma(R,\theta)$ is the number of clusters, $k$ is the normalization constant, $\theta$ is the $PA$ measured counterclockwise from the north, and $\alpha$ is the value of the power-law exponent in the surface density fit. The values obtained from these fits are listed in Table \ref{Tabla_distr_espacial}. Due to the low number of objects presented by the GC subpopulation with redder colours (group D), we did not perform the fit on this sample. 
Regarding this last group, since the galaxy does not seem to show the clear existence of dust that could turn the typical red GCs even redder in colour, the presence of the group D could be associated with those objects called diffuse star clusters \citep[e.g.,][]{Larsen2000,Forbes2014,Escudero2020} of which NGC\,4382 would present an excess \citep{Peng2006b,Liu2016}. 

\begin{table}
\centering
\scriptsize
\begin{tabular}{lccccc}
\multicolumn{6}{c}{}\\
\hline
\hline
\multicolumn{1}{c}{\textbf{Population}} &
\multicolumn{1}{c}{\textbf{Slope ({\it b})}} &
\multicolumn{1}{c}{\textbf{Zero point ({\it a})}} &
\multicolumn{1}{c}{\textbf{$\chi^2_\nu$}} & 
\multicolumn{1}{c}{\textbf{$\epsilon$}} &
\multicolumn{1}{c}{\textbf{$PA$\,(degrees)}} \\
\hline
All        & -1.39$\pm$0.11  & ~~0.29$\pm$0.23  & 1.7  & 0.19$\pm$0.12 & 91$\pm$22 \\

Blue (A)   & -0.88$\pm$0.26  & -1.31$\pm$0.54  & 1.7  & 0.27$\pm$0.29 & 87$\pm$35 \\

Interm. (B) & -1.53$\pm$0.28 & ~~0.13$\pm$0.55 & 2.0  & 0.16$\pm$0.18 & 38$\pm$30 \\

Red (C)    & -1.51$\pm$0.16  & -0.06$\pm$0.32  & 1.1  & 0.32$\pm$0.12 & 89$\pm$11 \\

+Red (D)   & -1.06$\pm$0.14  & -1.19$\pm$0.26  & 0.5  &    ---      & --- \\

\hline
\end{tabular}
\caption{Fitted parameters of slope ({\it b}) and zero point ({\it a}) corresponding to the considered power-law function. The fourth column indicates the values of reduced chi-square obtained from the fit of this function. The last two columns indicate the ellipticity ($\epsilon$) and position angle ($PA$) values obtained for the GC system and for each subpopulation in NGC\,4382. 
}
\label{Tabla_distr_espacial}
\end{table}

It should be noted that the results listed in Table \ref{Tabla_distr_espacial} show high uncertainties in the parameters obtained mainly in groups A and B. In this sense, the only subpopulation that presents a significant elongation is group C. On the other hand, for the intermediate subpopulation (group B)  a significant overdensity of objects is observed towards $PA=210$ degrees.
This characteristic is in agreement with that obtained by \citet{DAbrusco2015} where they determined several overdensity structures around NGC\,4382.
For its part, although red GCs generally resemble the shape of the host galaxy as observed in other early-type galaxies \citep{Kartha2014,Escudero2018}, the GC system of NGC\,4382 presents a lower elongation value compared to the red candidates (group C). This difference is probably due to the contribution of the intermediate group of clusters that smoothes the overall shape of the system. 
However, \citet{Cortesi2016} find that the red GC system of NGC\,1023 is more elongated than the blue GC system, resembling the disk surface brightness profile.

Comparing the ellipticity and the position angle obtained here for the GC system with the values of K19
at the same galactocentric radius ($R_\mathrm{gal}<2.5$ arcmin; 13.1 kpc), a good agreement is obtained on the ellipticity ($\epsilon\sim0.2$), while the $PA$ obtained by these authors is smaller ($\sim60$ degrees) compared to that obtained in this work (Table \ref{Tabla_distr_espacial}).


\section{Spectroscopic Analysis}
\label{spectr_analysis}
\subsection{Radial Velocities}
\label{rad_vel}
As mentioned in Section \ref{spect_data}, we analyse the spectra of 53 objects corresponding to our Gemini programmes GN-2016A-Q-62 and the programme GN-2015A-Q-207 of K18
(28 and 25 objects, respectively). The determination of the radial velocities of both datasets was carried out using the {\sc{fxcor}} task in the {\sc{rv}} package of {\sc{iraf}}. {\sc{fxcor}} uses the cross-correlation method \citep{Tonry1979} by comparing a spectrum whose radial velocity and dispersion velocity are unknown, with template spectra. In this work, we used as templates the stellar population synthesis models MILES \citep{Vazdekis2010}, considering the grid of models of $-2.27<\mathrm{[Z/H]}<+0.4$ dex, $1.0<\mathrm{age}<14.0$ Gyr, [$\alpha$/Fe]$=0$ and 0.4, and with a bimodal IMF of slope 1.3. The resolution of this dataset is 2.51 \AA, superior to our GMOS data. 
The final radial velocities for each object were obtained by applying a 3$\sigma$ clipped mean of the measured values. In addition, the radial velocity errors were obtained as the mean of errors measured by {\sc{fxcor}}. Since the typical average velocity dispersion for GCs \citep[e.g.,][]{Baumgardt2018} is below the resolution in our instrumental configuration, we were unable to estimate velocity dispersion in our dataset.

\begin{figure}
	\includegraphics[width=0.99\columnwidth]{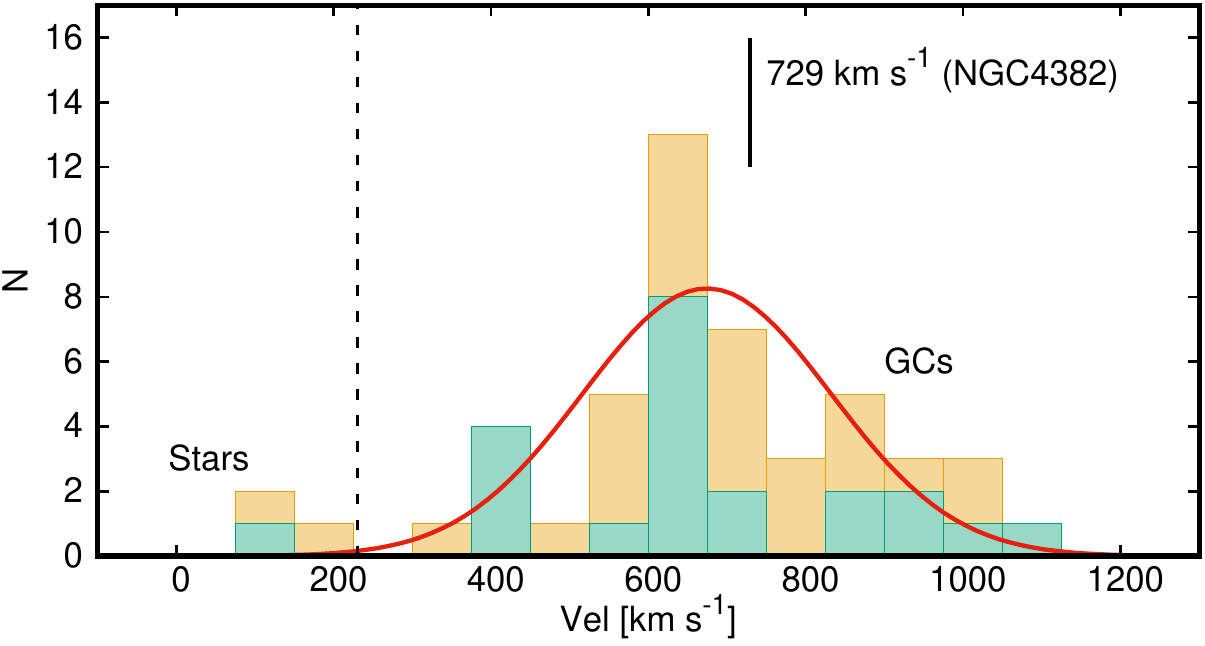}
    \caption{Stacked histogram of the radial velocity of objects belonging to the programme GN-2016A-Q-62 (orange histogram) and programme GN-2015A-Q-207 (green histogram). 
    The red line indicates the Gaussian fit performed on the sum of both samples. Vertical dashed line indicates the separation boundary between GCs associated with the galaxy and MW stars. The continuous vertical line indicates the systemic velocity of NGC\,4382 ($V_\mathrm{sys}$=729 km\,s$^{-1}$).}
    \label{fig:hist_velocity}
\end{figure}

Figure \ref{fig:hist_velocity} shows the velocity distribution of the analysed objects, together with the Gaussian fit obtained for the dataset ($\mu=674\pm35$ km\,s$^{-1}$ and $\sigma=158\pm36$ km\,s$^{-1}$). We consider the objects associated with the galaxy \citep[$V_\mathrm{sys}=729\pm2$ km\,s$^{-1}$;][]{Smith2000} with radial velocities within $\mu \pm 3\sigma$. Of the 28 objects in our mask, 25 were confirmed as GC associated with NGC\,4382. The remaining 3 objects correspond to two MW stars and possibly a background galaxy ($z\sim0.03$). These three objects were discarded for the following analyses in this work. Regarding the GCs confirmed by K18,
they were reconfirmed in our analysis. Table \ref{Tab_vr} lists the values obtained for each fitted object in both samples.

\begin{landscape}
\begin{table}
\centering
\scriptsize
\begin{tabular}{lcccccccccccc}
\hline
\hline
\multicolumn{1}{c}{\textbf{ID}} &
\multicolumn{1}{c}{\textbf{$\alpha$ (J2000)}} &
\multicolumn{1}{c}{\textbf{$\delta$ (J2000)}} &
\multicolumn{1}{c}{\textbf{$g'_0$}} &
\multicolumn{1}{c}{\textbf{$r'_0$}} &
\multicolumn{1}{c}{\textbf{$i'_0$}} &
\multicolumn{1}{c}{\textbf{Vel}} &
\multicolumn{1}{c}{\textbf{S/N (5800\,\AA)}} &
\multicolumn{1}{c}{\textbf{$(g'-r')_0$}} &
\multicolumn{1}{c}{\textbf{$(g'-i')_0$}} &
\multicolumn{1}{c}{\textbf{Age}} &
\multicolumn{1}{c}{\textbf{[Fe/H]}} &
\multicolumn{1}{c}{\textbf{$[\alpha/\mathrm{Fe}]$}} \\
\multicolumn{1}{c}{} &
\multicolumn{1}{c}{(deg)} &
\multicolumn{1}{c}{(deg)} &
\multicolumn{1}{c}{(mag)} &
\multicolumn{1}{c}{(mag)} &
\multicolumn{1}{c}{(mag)} &
\multicolumn{1}{c}{(km\,s$^{-1}$)} &
\multicolumn{1}{c}{per\,\AA} &
\multicolumn{1}{c}{(mag)} &
\multicolumn{1}{c}{(mag)} &
\multicolumn{1}{c}{(Gyr)} &
\multicolumn{1}{c}{(dex)} &
\multicolumn{1}{c}{(dex)} \\
\hline
4   & 186.30179 & 18.240056 & 22.314$\pm$0.009 & ----             & 21.504$\pm$0.018 & 997$\pm$25 & 17.6   & ---             & 0.810$\pm$0.020 & 10.9$\pm$3.8 & -1.46$\pm$0.13 & 0.00$\pm$0.20 \\
8   & 186.33221 & 18.252778 & 22.609$\pm$0.029 & ----             & 21.804$\pm$0.041 & 598$\pm$29 & 17.7   & ---             & 0.805$\pm$0.050 & 13.1$\pm$1.6 & -1.73$\pm$0.20 & 0.39$\pm$0.16 \\
19  & 186.33721 & 18.242528 & 23.012$\pm$0.028 & ----             & 22.194$\pm$0.036 & 967$\pm$20 & 13.8   & ---             & 0.818$\pm$0.045 & 6.8$\pm$2.4  & -1.51$\pm$0.21 & 0.40$\pm$0.30 \\
20  & 186.34246 & 18.244083 & 23.026$\pm$0.020 & ----             & 22.149$\pm$0.023 & 542$\pm$28 & 12.9   & ---             & 0.877$\pm$0.030 & 9.4$\pm$4.5  & -1.65$\pm$0.26 & 0.39$\pm$0.29 \\
22  & 186.36871 & 18.265528 & 23.215$\pm$0.014 & ----             & 22.435$\pm$0.010 & 801$\pm$40 &  8.1   & ---             & 0.780$\pm$0.017 & ---          & ---            & ----          \\
23  & 186.35029 & 18.245917 & 23.270$\pm$0.012 & ----             & 22.369$\pm$0.018 & 700$\pm$31 & 10.8   & ---             & 0.901$\pm$0.021 & 1.8$\pm$0.7  & -0.18$\pm$0.20 & 0.15$\pm$0.22 \\
25  & 186.32908 & 18.221667 & 22.158$\pm$0.015 & 21.649$\pm$0.010 & 21.306$\pm$0.018 & 843$\pm$24 & 20.1   & 0.509$\pm$0.018 & 0.852$\pm$0.023 & 9.8$\pm$4.4  & -1.48$\pm$0.14 & 0.10$\pm$0.40 \\
27  & 186.37596 & 18.263500 & 22.359$\pm$0.020 & ----             & 21.513$\pm$0.019 & 665$\pm$27 & 17.7   & ---             & 0.846$\pm$0.027 & 14.0$\pm$1.8 & -1.77$\pm$0.22 & 0.50$\pm$0.22 \\
64  & 186.37283 & 18.214417 & 22.434$\pm$0.009 & 21.866$\pm$0.011 & 21.391$\pm$0.012 & 639$\pm$12 & 15.1   & 0.568$\pm$0.014 & 1.043$\pm$0.015 & 2.2$\pm$0.4  & -0.34$\pm$0.17 & 0.11$\pm$0.15 \\
90  & 186.35996 & 18.212361 & 21.315$\pm$0.011 & ----             & 20.293$\pm$0.012 & 815$\pm$15 & 29.5   & ---             & 1.022$\pm$0.016 & 2.9$\pm$0.8  & -0.59$\pm$0.07 & 0.26$\pm$0.08 \\
105 & 186.35637 & 18.222861 & 23.218$\pm$0.018 & 22.728$\pm$0.016 & 22.434$\pm$0.016 & 527$\pm$40 & 10.1   & 0.490$\pm$0.024 & 0.784$\pm$0.024 & 6.3$\pm$2.9  & -1.55$\pm$0.50 & 0.00$\pm$0.20 \\
126 & 186.34829 & 18.221611 & 23.042$\pm$0.021 & 22.407$\pm$0.016 & 22.006$\pm$0.022 & 852$\pm$18 & 12.7   & 0.635$\pm$0.026 & 1.036$\pm$0.030 & 1.4$\pm$0.4  & ~0.02$\pm$0.09 & 0.09$\pm$0.18 \\
133 & 186.35429 & 18.239222 & 22.831$\pm$0.015 & ----             & 22.010$\pm$0.020 & 500$\pm$35 & 13.5   & ---             & 0.821$\pm$0.025 & 13.8$\pm$5.7 & -1.72$\pm$0.26 & 0.11$\pm$0.33 \\
165 & 186.37158 & 18.210639 & 22.275$\pm$0.007 & 21.710$\pm$0.009 & 21.109$\pm$0.007 & 746$\pm$16 & 18.8   & 0.565$\pm$0.011 & 1.166$\pm$0.009 & 10.2$\pm$3.4 & -0.41$\pm$0.12 & 0.37$\pm$0.10 \\
166 & 186.33346 & 18.213639 & 22.252$\pm$0.012 & 21.539$\pm$0.009 & 21.021$\pm$0.010 & 610$\pm$13 & 19.8   & 0.713$\pm$0.015 & 1.231$\pm$0.015 & 13.8$\pm$2.2 & -0.54$\pm$0.06 & 0.22$\pm$0.09 \\
167 & 186.34233 & 18.212472 & 21.845$\pm$0.009 & 21.225$\pm$0.006 & 20.809$\pm$0.010 & 620$\pm$10 & 23.3   & 0.620$\pm$0.010 & 1.036$\pm$0.013 & 13.4$\pm$5.3 & -0.63$\pm$0.08 & 0.31$\pm$0.08 \\
168 & 186.36442 & 18.209222 & 21.548$\pm$0.020 & 21.124$\pm$0.010 & 20.723$\pm$0.021 & 360$\pm$23 & 25.8   & 0.424$\pm$0.022 & 0.825$\pm$0.029 & 11.5$\pm$3.2 & -1.44$\pm$0.12 & 0.31$\pm$0.10 \\
172 & 186.33137 & 18.208611 & 22.360$\pm$0.011 & 21.703$\pm$0.008 & 21.218$\pm$0.008 & 828$\pm$11 & 17.8   & 0.657$\pm$0.013 & 1.142$\pm$0.013 & 11.4$\pm$3.5 & -0.41$\pm$0.15 & 0.28$\pm$0.12 \\ 
173 & 186.34187 & 18.205222 & 22.644$\pm$0.012 & 22.140$\pm$0.015 & 21.737$\pm$0.012 & 757$\pm$36 & 14.2   & 0.504$\pm$0.019 & 0.907$\pm$0.016 & 9.6$\pm$3.7  & -1.31$\pm$0.11 & 0.40$\pm$0.22 \\
175 & 186.35392 & 18.201500 & 21.871$\pm$0.016 & 21.384$\pm$0.011 & 20.922$\pm$0.014 & 688$\pm$15 & 20.2   & 0.487$\pm$0.019 & 0.949$\pm$0.021 & 13.0$\pm$4.4 & -0.96$\pm$0.11 & 0.28$\pm$0.15 \\
176 & 186.34504 & 18.201028 & 22.996$\pm$0.015 & 22.418$\pm$0.021 & 21.983$\pm$0.016 & 725$\pm$33 &  9.5   & 0.578$\pm$0.025 & 1.013$\pm$0.022 & ---          &  ---           &  ---          \\
179 & 186.34867 & 18.198889 & 22.864$\pm$0.028 & 22.443$\pm$0.024 & 22.093$\pm$0.021 & 640$\pm$22 &  5.7   & 0.421$\pm$0.036 & 0.771$\pm$0.035 & ---          &  ---           &  ---          \\
181 & 186.33904 & 18.198222 & 22.630$\pm$0.012 & 22.119$\pm$0.015 & 21.641$\pm$0.009 & 983$\pm$40 & 12.8   & 0.511$\pm$0.019 & 0.989$\pm$0.015 & 3.6$\pm$1.8  & -0.45$\pm$0.20 & 0.16$\pm$0.22 \\
184 & 186.35904 & 18.192833 & 22.465$\pm$0.018 & 22.007$\pm$0.016 & 21.520$\pm$0.010 & 709$\pm$24 & 14.1   & 0.458$\pm$0.024 & 0.945$\pm$0.020 & 14.0$\pm$2.1 & -1.05$\pm$0.17 & 0.29$\pm$0.27 \\
186 & 186.35571 & 18.192861 & 21.784$\pm$0.013 & 21.281$\pm$0.009 & 20.687$\pm$0.012 & 554$\pm$16 & 21.6   & 0.503$\pm$0.015 & 1.097$\pm$0.017 & 11.5$\pm$3.2 & -0.63$\pm$0.08 & -0.01$\pm$0.13 \\
Star/Gal &      &           &                  &                  &                 &             &        &                 &                 &              &                &              \\ 
2   & 186.31875 & 18.271528 & 22.221$\pm$0.010 & ----             & 21.358$\pm$0.008 & 103$\pm$17 & 18.1   &     ----        &     ----        &     ----     &    ----        &    ----       \\ 
3   & 186.34425 & 18.287667 & 22.465$\pm$0.008 & ----             & 21.728$\pm$0.010 & 165$\pm$35 & 16.5   &     ----        &     ----        &     ----     &    ----        &    ----       \\ 
5  & 186.31237 & 18.243833 & 22.583$\pm$0.045 & ----              & 21.501$\pm$0.066 & $z$=0.03   & 15.1   &     ----        &     ----        &     ----     &    ----        &    ----       \\ 
\hline
21  & 186.30570 & 18.16026 & 22.090$\pm$0.014 & 21.502$\pm$0.010 & 20.896$\pm$0.011 & 1038$\pm$14 & 21.9   & 0.588$\pm$0.017 & 1.194$\pm$0.017 & 3.8$\pm$1.3  & ~0.13$\pm$0.07 & 0.27$\pm$0.07 \\
48  & 186.30765 & 18.20636 & 22.115$\pm$0.009 & 21.600$\pm$0.007 & 21.258$\pm$0.004 & 971$\pm$26 & 20.9    & 0.515$\pm$0.011 & 0.857$\pm$0.009 & 7.9$\pm$3.0  & -0.79$\pm$0.11 & 0.04$\pm$0.12 \\
39  & 186.32040 & 18.16042 & 22.047$\pm$0.010 & 21.619$\pm$0.007 & 21.114$\pm$0.005 & 691$\pm$12 & 18.7    & 0.428$\pm$0.012 & 0.933$\pm$0.011 & 13.8$\pm$2.0 & -0.98$\pm$0.12 & 0.34$\pm$0.13 \\
45  & 186.32670 & 18.19356 & 22.027$\pm$0.014 & 21.533$\pm$0.006 & 21.127$\pm$0.006 & 1067$\pm$24 & 21.7   & 0.494$\pm$0.015 & 0.900$\pm$0.015 & 5.4$\pm$1.4  & -1.18$\pm$0.12 & 0.44$\pm$0.14 \\
20  & 186.32940 & 18.15972 & 21.573$\pm$0.011 & 21.147$\pm$0.008 & 20.667$\pm$0.010 & 721$\pm$12 & 25.1    & 0.426$\pm$0.013 & 0.906$\pm$0.014 & 2.1$\pm$0.3  & ~0.17$\pm$0.08 & 0.00$\pm$0.06 \\
 5  & 186.33330 & 18.19563 & 21.515$\pm$0.010 & 20.936$\pm$0.005 & 20.464$\pm$0.005 & 930$\pm$24 & 27.9    & 0.579$\pm$0.011 & 1.051$\pm$0.011 & 12.0$\pm$2.6 & -0.72$\pm$0.06 & 0.31$\pm$0.06 \\
29  & 186.33675 & 18.18428 & 21.831$\pm$0.018 & 21.347$\pm$0.008 & 20.919$\pm$0.010 & 602$\pm$18 & 23.1    & 0.484$\pm$0.019 & 0.912$\pm$0.020 & 13.8$\pm$2.2 & -1.08$\pm$0.11 & 0.22$\pm$0.15 \\
 6  & 186.34080 & 18.19698 & 21.194$\pm$0.010 & 20.625$\pm$0.004 & 20.155$\pm$0.003 & 636$\pm$7  & 34.7    & 0.569$\pm$0.010 & 1.039$\pm$0.010 & 1.5$\pm$0.2  & ~0.18$\pm$0.04 & 0.00$\pm$0.06 \\
25  & 186.34845 & 18.17561 & 21.836$\pm$0.014 & 21.329$\pm$0.008 & 20.876$\pm$0.009 & 872$\pm$20 & 17.1    & 0.507$\pm$0.016 & 0.960$\pm$0.016 & 14.0$\pm$2.8 & -0.38$\pm$0.11 & 0.33$\pm$0.10 \\
 7  & 186.35250 & 18.19900 & 20.403$\pm$0.006 & 19.873$\pm$0.003 & 19.408$\pm$0.003 & 615$\pm$12 & 51.3    & 0.530$\pm$0.006 & 0.995$\pm$0.006 & 2.0$\pm$0.1  & ~0.19$\pm$0.02 & 0.04$\pm$0.03 \\
 2  & 186.35415 & 18.18246 & 21.569$\pm$0.015 & 21.002$\pm$0.008 & 20.488$\pm$0.008 & 614$\pm$8  & 30.2    & 0.567$\pm$0.017 & 1.081$\pm$0.017 & 14.0$\pm$2.6 & -0.33$\pm$0.07 & 0.20$\pm$0.05 \\
23  & 186.35625 & 18.17223 & 21.382$\pm$0.010 & 20.944$\pm$0.007 & 20.581$\pm$0.013 & 446$\pm$18 & 28.6    & 0.438$\pm$0.012 & 0.801$\pm$0.016 & 12.4$\pm$2.9 & -1.45$\pm$0.08 & 0.17$\pm$0.09 \\
33  & 186.36075 & 18.20630 & 21.790$\pm$0.013 & 21.405$\pm$0.013 & 20.887$\pm$0.007 & 449$\pm$14 & 26.1    & 0.385$\pm$0.018 & 0.903$\pm$0.014 & 11.6$\pm$2.5 & -1.26$\pm$0.10 & 0.22$\pm$0.13 \\
30  & 186.36570 & 18.18630 & 21.688$\pm$0.016 & 21.172$\pm$0.006 & 20.687$\pm$0.006 & 669$\pm$8  & 25.9    & 0.516$\pm$0.017 & 1.001$\pm$0.017 & 11.2$\pm$2.9 & -0.90$\pm$0.07 & 0.23$\pm$0.07 \\
 4  & 186.36885 & 18.18572 & 20.947$\pm$0.004 & 20.517$\pm$0.004 & 20.109$\pm$0.012 & 865$\pm$12 & 37.3    & 0.430$\pm$0.005 & 0.838$\pm$0.012 & 13.8$\pm$1.3 & -1.51$\pm$0.06 & 0.22$\pm$0.08 \\
32  & 186.37335 & 18.19373 & 21.659$\pm$0.008 & 21.253$\pm$0.011 & 20.757$\pm$0.018 & 642$\pm$10 & 26.6    & 0.406$\pm$0.013 & 0.902$\pm$0.019 & 1.3$\pm$0.2  & ~0.14$\pm$0.04 & 0.02$\pm$0.08 \\
14  & 186.35340 & 18.20596 & 22.232$\pm$0.010 & 21.705$\pm$0.007 & 21.084$\pm$0.003 & 416$\pm$14 & 20.7    & 0.527$\pm$0.012 & 1.148$\pm$0.010 & 6.1$\pm$1.8  & -0.25$\pm$0.08 & 0.20$\pm$0.08 \\
43  & 186.38730 & 18.18649 & 22.308$\pm$0.012 & 21.785$\pm$0.007 & 21.229$\pm$0.005 & 536$\pm$12 & 19.4    & 0.523$\pm$0.013 & 1.079$\pm$0.013 & 14.0$\pm$1.5 & -0.56$\pm$0.14 & 0.30$\pm$0.14 \\
15  & 186.38940 & 18.20662 & 22.075$\pm$0.011 & 21.782$\pm$0.010 & 21.245$\pm$0.007 & 435$\pm$20 & 20.9    & 0.293$\pm$0.014 & 0.830$\pm$0.013 &  7.4$\pm$2.7 & -1.18$\pm$0.11 & 0.01$\pm$0.18 \\
40  & 186.39480 & 18.17053 & 22.490$\pm$0.010 & 21.805$\pm$0.007 & 21.411$\pm$0.006 & 620$\pm$18 & 18.0    & 0.685$\pm$0.012 & 1.079$\pm$0.011 & 12.7$\pm$3.9 & -0.39$\pm$0.14 & 0.23$\pm$0.14 \\
64* & 186.34515 & 18.18155 & 19.095$\pm$0.003 & 18.587$\pm$0.003 & 18.096$\pm$0.007 & 630$\pm$8  & 89.9    & 0.508$\pm$0.004 & 0.999$\pm$0.007 & 1.5$\pm$0.1  & ~0.23$\pm$0.01 & 0.18$\pm$0.02 \\
61** & 186.35040 & 18.19105 & ----            & ----             & ----             & 738$\pm$6  & 645.    & ----            & ----            & 2.5$\pm$0.1  & ~0.29$\pm$0.01 & 0.13$\pm$0.01 \\
Star/Gal &      &          &                  &                  &                  &            &         &                 &                 &              &                &               \\
12  & 186.32385 & 18.17269 & 21.810$\pm$0.011 & 21.473$\pm$0.008 & 21.115$\pm$0.009 & 115$\pm$12 & 21.6    &    ----         &    ----         &     ----     &    ----        &    ----       \\
\hline
\end{tabular}
\caption{Spectrophotometric parameters of the objects associated with our Gemini programme GN-2016A-Q-62 and the programme GN-2015A-Q-207 (PI: Myung Gyoon Lee), derived in this work. The second half of the table corresponds to this last programme. * and ** indicate the hypercompact cluster M85-HCC1 \citep{Sandoval2015} and the nucleus of NGC\,4382, respectively.}
\label{Tab_vr}
\end{table}
\end{landscape}


\subsection{Full Spectral Fitting}
\label{full_spectral}

The analysis of the stellar populations was carried out using the full spectral fitting technique through the ULySS \citep[University of Lyon Spectroscopic Analysis Software;][]{Koleva2009} algorithm. ULySS fits synthetic single stellar population (SSP) models to the observed spectra, providing SSP equivalent stellar population parameters. In this work, we used the PEGASE-HR model grid \citep{Leborgne2004} resolved in [$\alpha$/Fe] \citep{Prugniel2012}. Assuming the initial mass function of Salpeter \citep[IMF;][]{Salpeter1955} and the ELODIE3.2 stellar library \citep{Prugniel2001,Wu2011}, these models cover a wavelength range of 3900-6800 \AA\, with a resolution of FWHM=0.55 \AA. These synthetic SSP models have ages between 0.001 and 20 Gyr, metallicities in the range of $-2.3<\mathrm{[Fe/H]}<+0.69$ dex and [$\alpha$/Fe]=0 and 0.4 dex. In this work, we carry out the spectral fits considering the models in the age range 0.001 to 14 Gyr. 

The fits made with ULySS were carried out using the largest possible spectral range according to each science spectrum ($\sim4000-6700$\,\AA) since the precision of the results mostly depends on the total S/N ratio \citep{Koleva2008}. In this process, a multiplicative polynomial of order 20 was considered to absorb possible errors in flux calibration. Furthermore, we consider the parameter {\sc{clean}} of ULySS to exclude outliers in our spectra due to possible residues by night sky emission and cosmic rays during the extraction process. Finally, we performed 200 Monte Carlo simulations to determine the standard deviations of the estimated luminosity-weighted stellar population parameters. Only in 3 objects in our sample, it was not possible to obtain their stellar parameters using UlySS. Different tests were carried out in order to obtain the convergence of the code. However, no positive results were obtained in the fits possibly due to the low S/N presented by these objects.
Table \ref{Tab_vr} lists the obtained values in this work. 

\begin{figure}
    \centering
	\includegraphics[width=0.8\columnwidth]{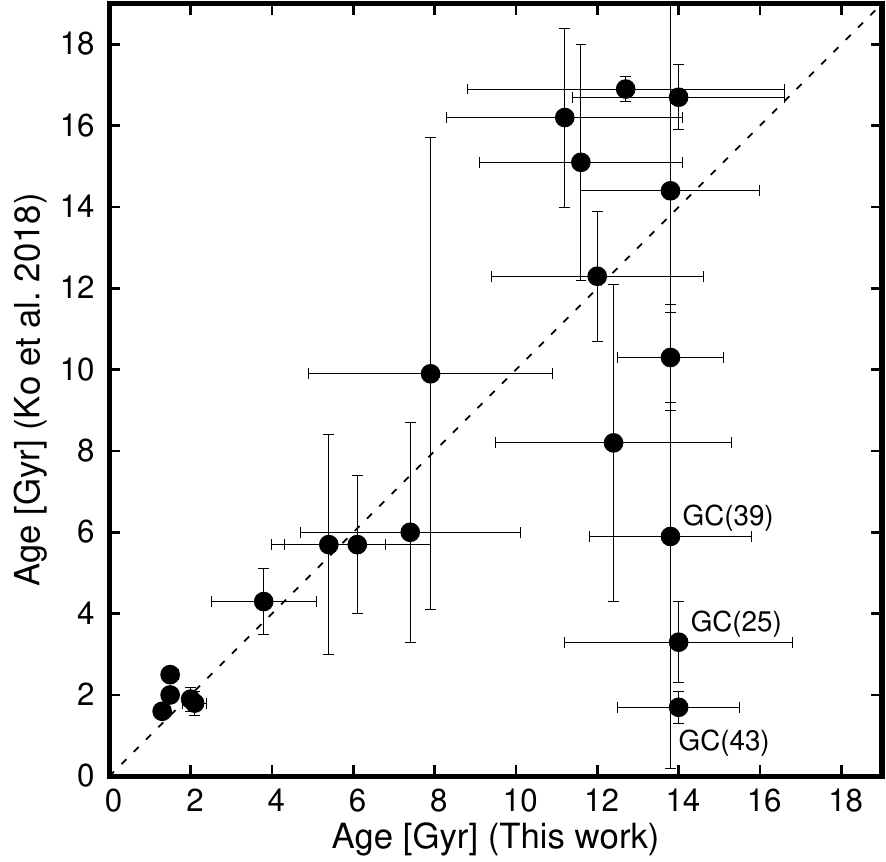}
	\includegraphics[width=0.84\columnwidth]{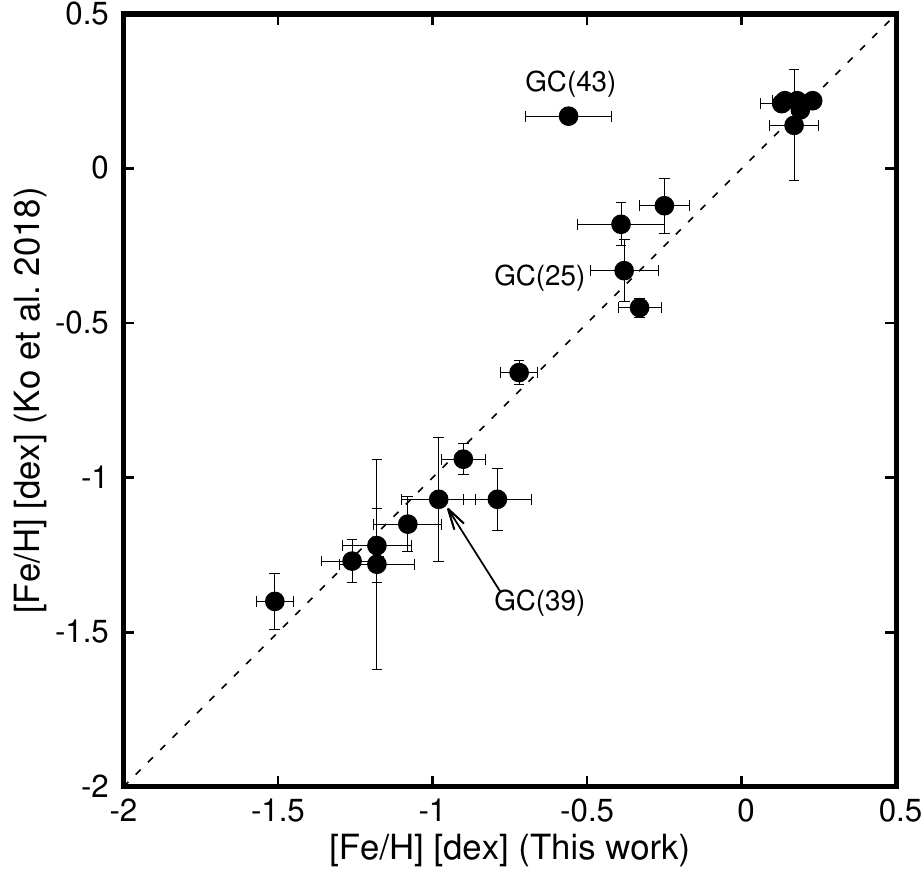}
	\includegraphics[width=0.84\columnwidth]{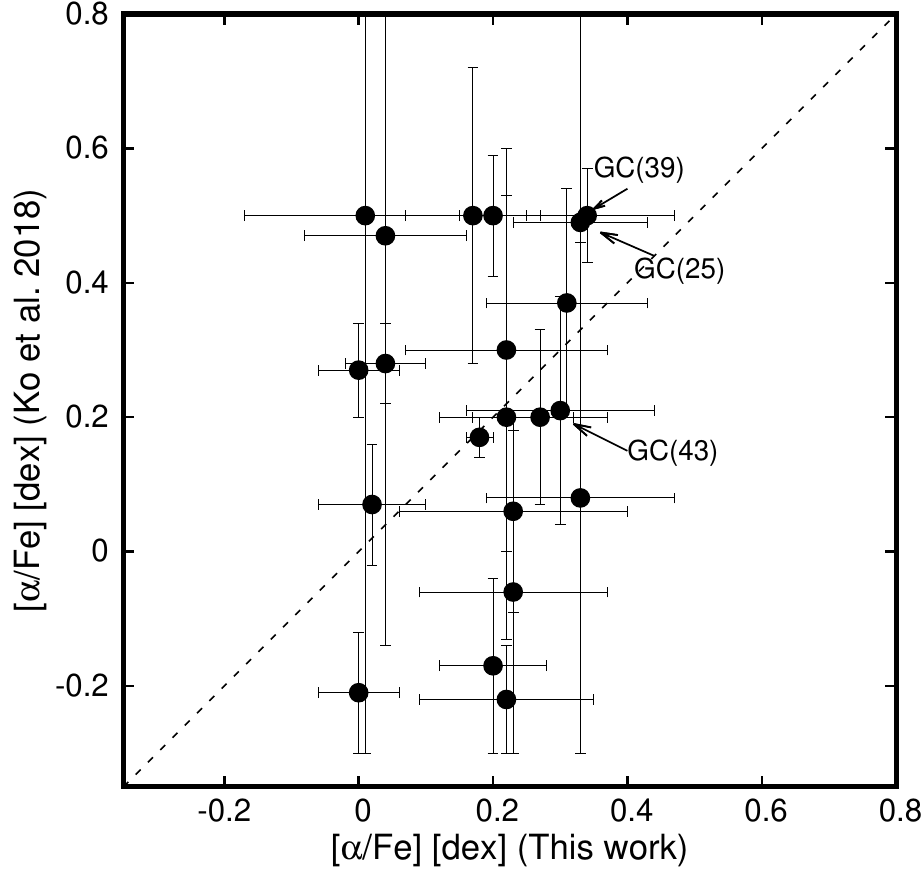}
    \caption{Comparison between the values of ages (top panel), metallicities (central panel) and $\alpha$-element abundances (bottom panel) of the objects studied in K18,
    and the values estimated here considering the SSP model resolved in [$\alpha$/Fe]. The [$\alpha$/Fe] values obtained by these authors were estimated using the Lick index method. 
    The dashed line indicates the one-to-one relation.}
    \label{fig:comparison}
\end{figure}

Although K18 also used ULySS in their analysis, the SSP model used by them did not consider the improvement in [$\alpha$/Fe]. The [$\alpha$/Fe] values derived in that work were obtained by a different technique, using Lick indices \citep{Burstein1984,Worthey1994,Worthey1997} and SSP models of \citet{Thomas2011}. In particular, those models have a wider range in [$\alpha$/Fe] ($-0.3$ to $+0.5$ dex) compared to the model used in this work. 
Therefore, in order to homogenize the results obtained between our sample of objects and those of K18, we recalculate the stellar population parameters of the latter considering the previously mentioned $\alpha$-resolved SSP model (Table \ref{Tab_vr}). 

We compare the values of age, metallicity and [$\alpha$/Fe] obtained by K18
with those calculated here. In Figure \ref{fig:comparison}, in general, the different parameters obtained with this new model are in good agreement within uncertainties with those of K18.
In the case of [$\alpha$/Fe], a significant dispersion can be seen when comparing the values recovered in this work with the ones obtained by K18.
However, it is necessary to emphasize that the values were obtained using different techniques and models. 
The top panel of Figure \ref{fig:comparison} shows three objects that stand out for clearly deviating from the one-to-one relationship, GC(43), GC(39) and GC(25). 
We carry out different tests on these objects, using shorter wavelength ranges, varying the multiplicative polynomial, without considering the parameter {\sc{clean}} of ULySS, and even reviewing the reduction process again. In all cases, we have not been able to find an explanation for such differences, since results similar to those shown in Table \ref{Tab_vr} were always obtained in the fits.

\subsection{Stellar Populations}
\label{populations}

As obtained in the photometric analysis of the colour distribution (see Section \ref{subpop}), as well as in other photometric and spectroscopic studies (e.g., K18, K19, K20),
the galaxy shows different GC subpopulations.
In particular, in K18
approximately one-third of their spectroscopic sample located within $R_\mathrm{gal}=3$ arcmin (15.6 kpc) have ages $<$6 Gyr ($\langle$age$\rangle$=$3.7\pm1.9$ Gyr), while the rest of the objects have old ages ($\langle$age$\rangle$=$13.3\pm3.3$ Gyr). 

\begin{figure}
    \includegraphics[width=0.99\columnwidth]{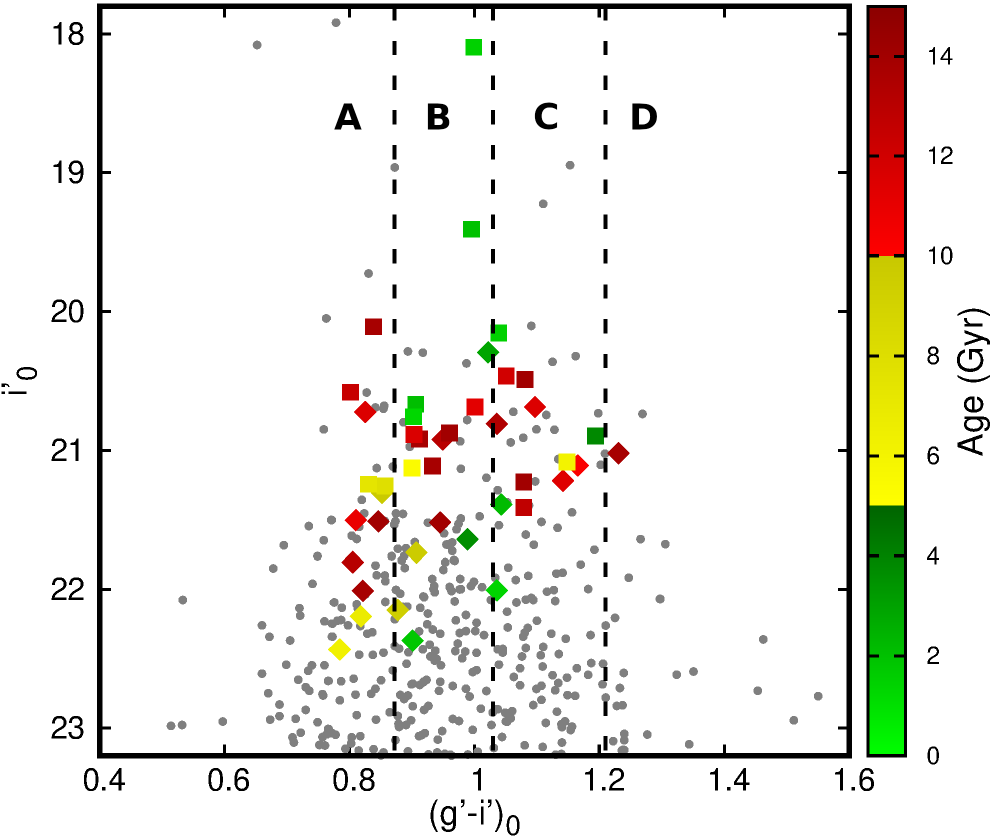}
	\includegraphics[width=0.99\columnwidth]{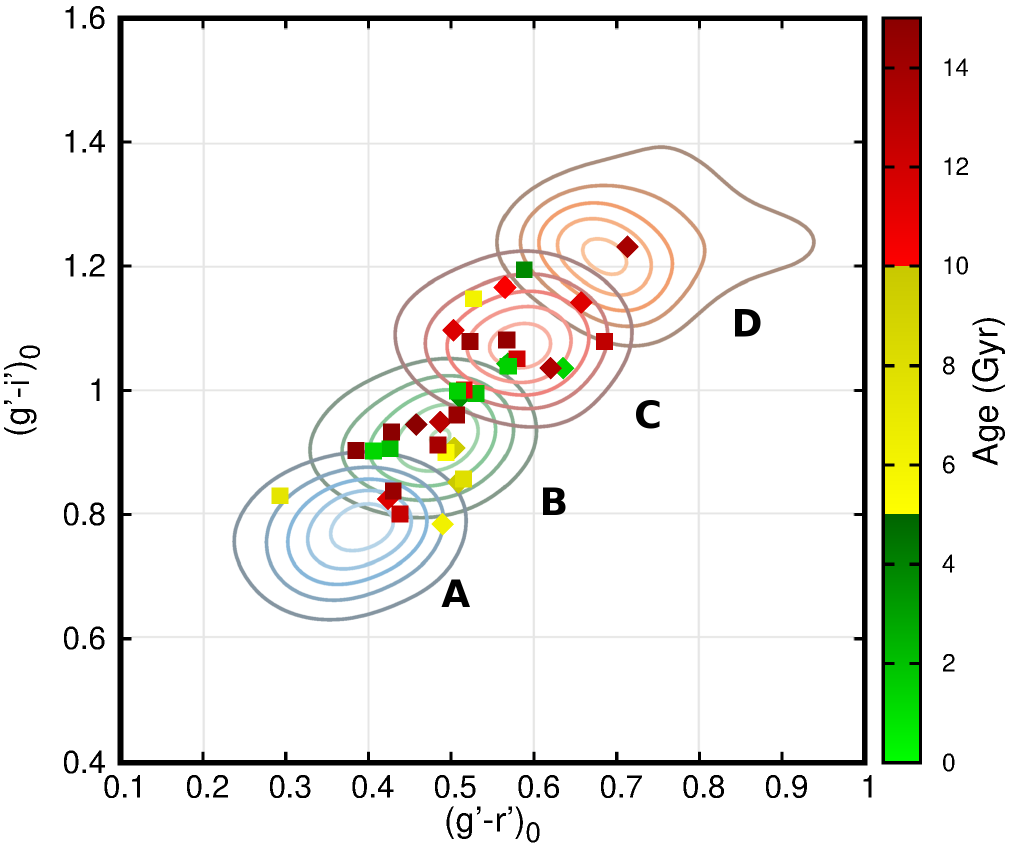}
    \caption{Upper panel: colour-magnitude diagram of the GC system of NGC4382 (grey points). Vertical dashed lines indicate the approximate separation in colour $(g'-i')_0$ of the different groups according to GMM. Bottom panel: colour-colour diagram as in Figure \ref{fig:AIC_BIC} indicating the four photometric GC groups identified by GMM (solid lines). Filled coloured rhombuses and squares indicate the GCs/HCC spectroscopically confirmed as a function of age corresponding to the programmes GN-2016A-Q-62 and GN-2015A-Q-207, respectively. Note that some GCs do not have colour $(g'-r')_0$ so they are not shown in the colour-colour diagram (see text).}
    \label{fig:col_mag_spec}
\end{figure}

Since in this work we added twice as many objects as the sample analyzed by K18
in the same region, and even some stellar population values estimated by these authors differ from those obtained here, we analyze the joint sample in order to have a more complete picture of the cluster formation and assembly events in the galaxy.
Figure \ref{fig:col_mag_spec} shows the colour-magnitude and colour-colour diagrams of the GC system of NGC\,4382 with the different groups identified by GMM (Section \ref{subpop}), and the location of the spectroscopically confirmed GCs. It should be noted that several of the GCs mainly associated with group A do not have information in colour $(g'-r')_0$, which is why there is a significant lack of objects observed in this region of the colour-colour diagram. 
As it can be seen in the figure, a good correlation is obtained between the photometric analysis and the spectroscopic results, that is, the GCs with ages less than 5 Gyr are associated with groups B and C, while the clusters with ages $>$5 Gyr present a wide range of colours associated with all the different groups. 
On the other hand, only one object (GC(166)) is found in the photometric region corresponding to group D associated with objects called diffuse star clusters (see Section \ref{spatial}). However, when observing its metallicity and brightness values (see Table \ref{Tab_vr}), they do not turn out to be typical values for this type of objects \citep[see][]{Peng2006b}, so it is probably a typical red GC.

\begin{figure}
	\includegraphics[width=0.99\columnwidth]{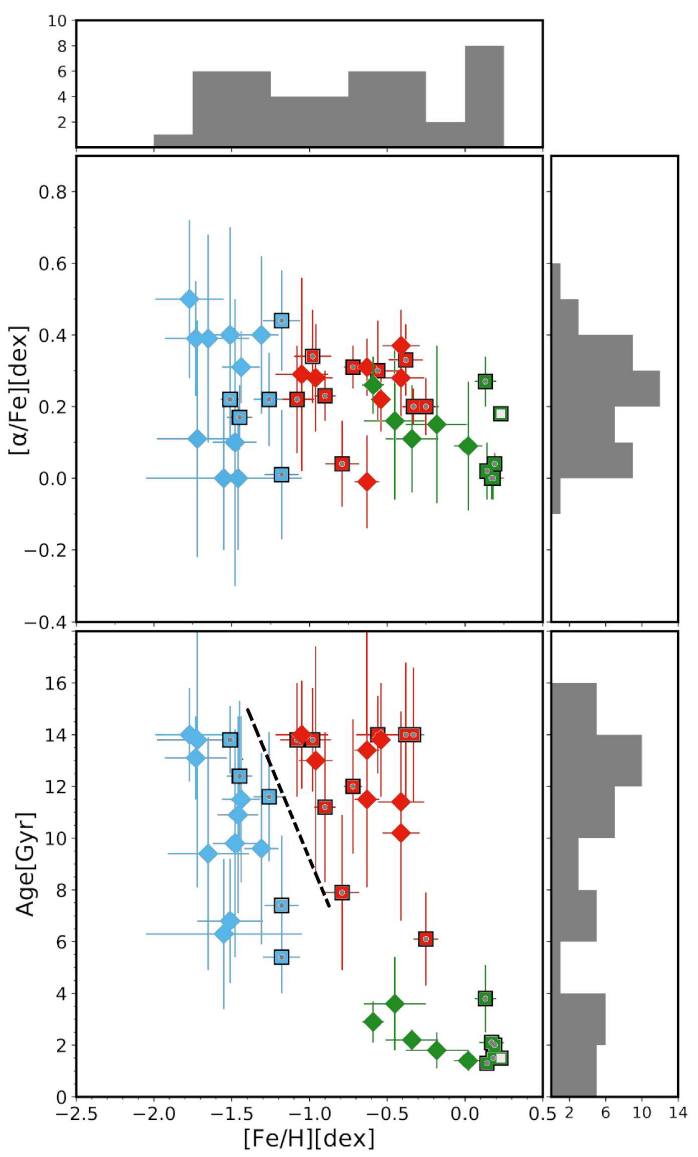}
    \caption{Diagrams [$\alpha$/Fe]-[Fe/H] and age-[Fe/H] of the objects associated with NGC\,4382. The rhombuses and squares correspond to the objects associated with the Gemini programmes GN-2016A-Q-62 and GN-2015A-Q-207, respectively. The squared filled square indicates the hypercompact cluster M85-HCC1. The blue, red, and green symbols indicate the old metal-poor, old metal-rich, and young GC subpopulations, respectively. The black dashed line in the lower panel indicates the limit used to separate the two groups of old GCs.}
    \label{fig:age_met_alfa}
\end{figure}

Figure \ref{fig:age_met_alfa} shows the relationships between the age, metallicity and $\alpha$-element abundance values of the objects in the sample obtained through the full spectral fitting technique. Based on the age distribution of the sample, it is observed in the lower panel of the figure the presence of a group of young objects ($<$5 Gyr; green symbols) with mean age, metallicity and $\alpha$-abundance of $2.2\pm0.9$ Gyr, $\mathrm{[Fe/H]}=-0.05\pm0.28$ dex and $\mathrm{[\alpha/Fe]}=0.11\pm0.10$ dex, respectively, stands out. Within this group, the hypercompact cluster M85-HCC1 is located (double square), which is considered a tidally stripped galaxy-centre according to \citet{Sandoval2015}. When comparing the aforementioned mean values for this young group with those of K18
($3.7\pm1.9$ Gyr, $\mathrm{[Fe/H]}=-0.26\pm0.62$ dex), a slight difference is obtained in age and metallicity, although within the uncertainties. Furthermore, it should be noted that the estimated values here are in good agreement with those of the stellar component of the nucleus of the galaxy (see Table \ref{Tab_vr}). 

On the other hand, objects with old ages ($>5$ Gyr) seem to show a small gap in the age-metallicity diagram (lower panel of Figure \ref{fig:age_met_alfa}). This bifurcated age-metallicity relationship, indicated with a dotted line in the figure, is similar to that observed for the halo and bulge-disc GCs of the MW \citep{Leaman2013}. Using this relationship, we separate the old GCs into two groups which present mean values $10.4\pm2.8$ Gyr, $\mathrm{[Fe/H]}=-1.48\pm0.18$ dex, $\mathrm{[\alpha/Fe]}=0.24\pm0.16$ dex (blue symbols) and $12.1\pm2.3$ Gyr, $\mathrm{[Fe/H]}=-0.64\pm0.26$ dex, $\mathrm{[\alpha/Fe]}=0.24\pm0.10$ dex (red symbols). These mean values are similar to those associated with the typical old metal-poor (blue) and old metal-rich (red) GC subpopulations \citep[e.g.,][]{Usher2012}, respectively, in early-type galaxies.

\begin{figure}
	\includegraphics[width=0.99\columnwidth]{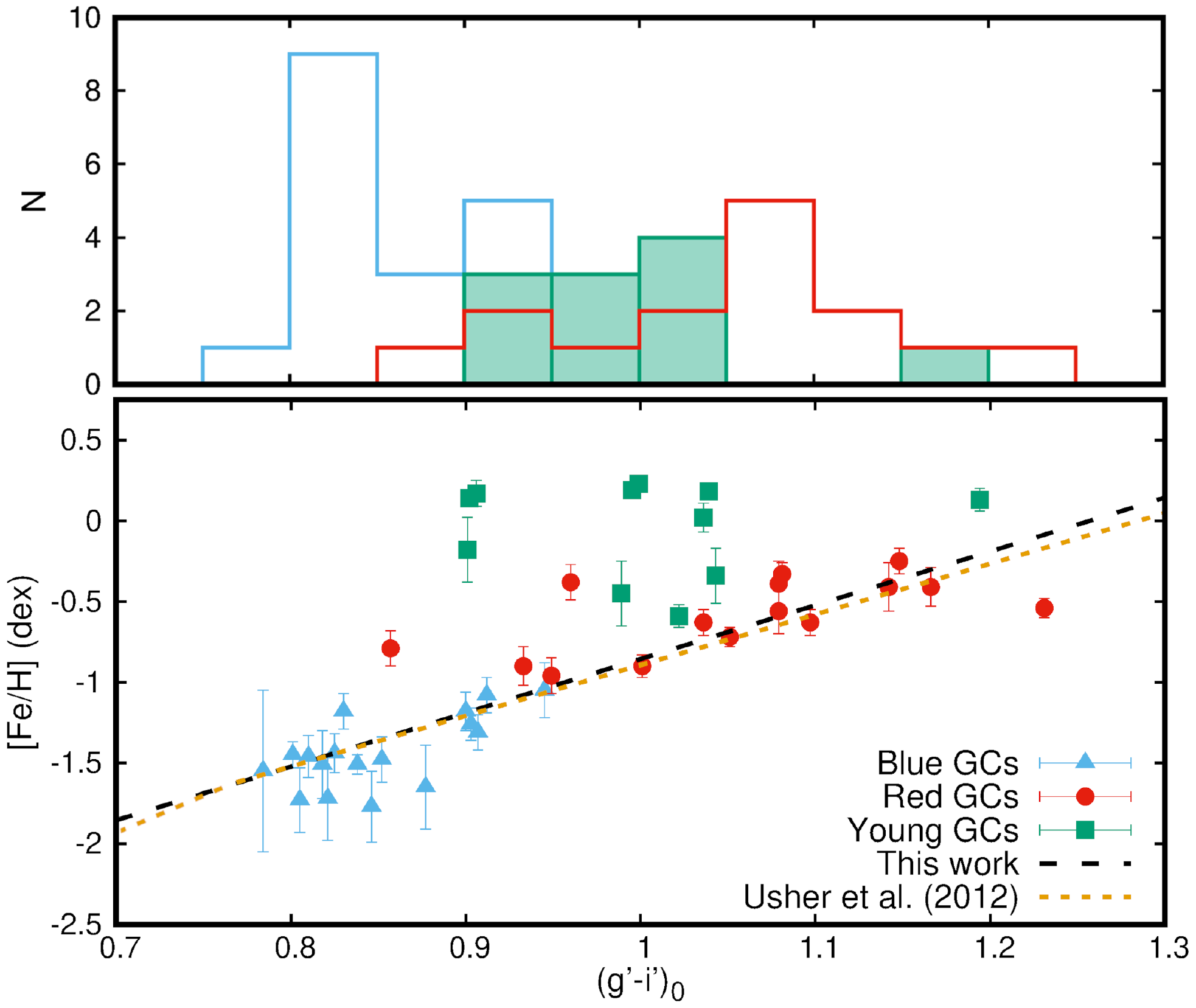}
    \caption{Colour-metallicity relation for the GCs of NGC\,4382. Blue triangles, red circles and green squares indicate the old metal-poor, old metal-rich and young GC subpopulations, respectively. Black dashed line and orange dotted line indicates the colour-metallicity relations obtained in this work and in \citet{Usher2012}, respectively. The upper panel shows the colour histograms of the different spectroscopic groups.}
    \label{fig:color_met}
\end{figure}

Considering the age and metallicity limits mentioned above to separate the typical old metal-poor and metal-rich clusters, and those young ones, we displayed the metallicity values as a function of the photometric colour $(g'-i')_0$ for these 3 groups (Figure \ref{fig:color_met}). The figure shows a clear trend followed by the old clusters, while the young ones seem to follow a different sequence. This colour-metallicity relationship is similar to those obtained for other early-type galaxies, particularly in the case of NGC\,1316 \citep{Sesto2018}, where its young GCs deviate from this relationship. In this work, we fit a linear relation 
considering only the old GCs, obtaining $\mathrm{[Fe/H]}=(3.33\pm0.32) \times (g'-i')_0 - (4.188\pm0.31)$. 
When comparing with the colour-metallicity relation obtained in \citet{Usher2012} for the colour values $(g'-i')_0>0.77$ mag (orange dotted line in Figure \ref{fig:color_met}), a good agreement is obtained between the two expressions. It is worth mentioning that the relation obtained in terms of total metallicity ([Z/H]) in \citet{Usher2012} was transformed to iron abundance ([Fe/H]) using expression 2 published in the same work. 

\begin{figure}
	\includegraphics[width=0.99\columnwidth]{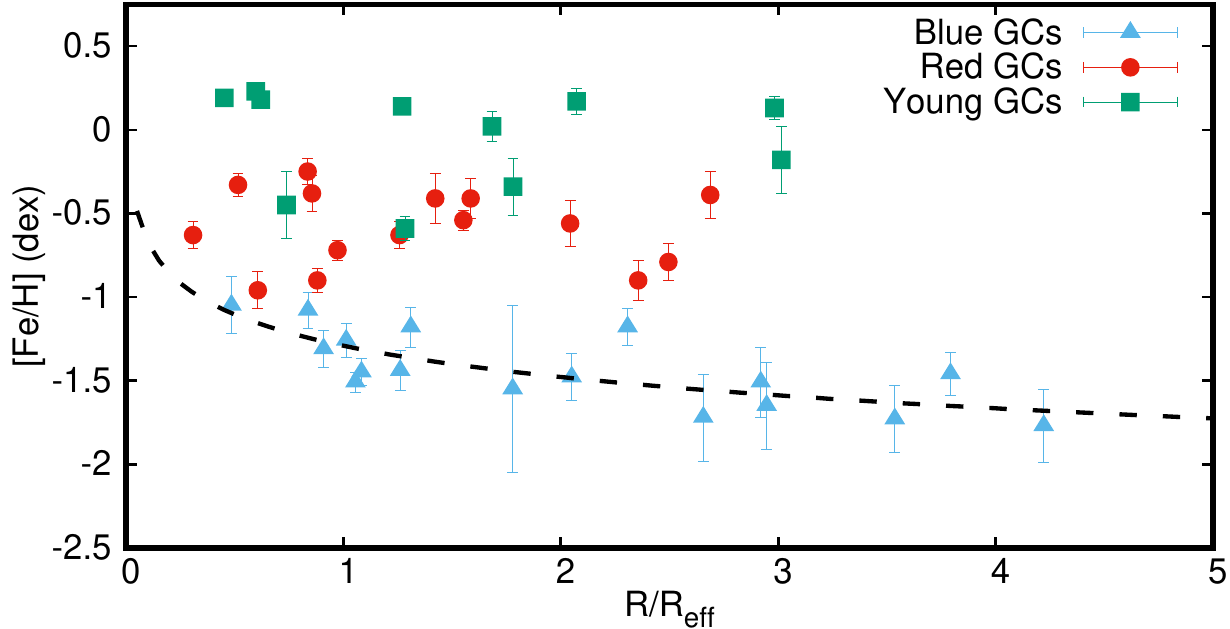}
    \caption{Radial metallicity distribution for the GCs of NGC\,4382. Blue triangles, red circles and green squares indicate the old metal-poor, old metal-rich and young GC subpopulations, respectively. Black dashed line indicates the metallicity gradient obtained for the old metal-poor GC subpopulation.}
    \label{fig:met_dist}
\end{figure}

On the other hand, we analyzed possible metallicity gradients in the GC subpopulations in order to identify different star formation processes \citep[e.g.,][]{Hopkins2009,Pipino2010} that could have occurred in NGC\,4382. Figure \ref{fig:met_dist} shows the radial metallicity distribution for the GCs of each subpopulation as a function of galactocentric distance. Both the old metal-rich group and the young group do not initially show a gradient along with the radial extent of the GC system, while the old metal-poor ones show a clear negative metallicity gradient. 
We fit the logarithmic relation $\mathrm{[Fe/H]} = a + b\,log(R/R_\mathrm{{eff}})$ for each subpopulation in the same radius range ($R/R_\mathrm{eff}<3.1$), where $R_\mathrm{eff}$ is the effective radius for NGC\,4382 \citep[$R_\mathrm{eff}=66$ arcsec;][]{Cappellari2011}. In the case of the old metal-rich and young clusters, we could not detect any significant metallicity gradient ($a=-0.60\pm0.15$, $b=-0.01\pm0.10$ and $a=-0.14\pm0.21$, $b=-0.07\pm0.14$), while for the old metal-poor clusters a clear relation is obtained ($a=-1.70\pm0.07$, $b=-0.27\pm0.06$). This last value is similar to those obtained in other early-type galaxies \citep[e.g.,][]{Forbes2011,Kartha2016}, which would indicate that part of the old metal-poor subpopulation would have been formed by dissipative collapse.
Alternatively, a similar gradient may be generated from a fairly sizable metal-rich accretion event whose GCs would be pulled closer to the centre because of the increased efficiency of dynamical friction as a function of mass \citep[e.g.,][]{Amorisco2019}. In the case of old metal-rich clusters, even though in general they have been found to show similar gradients to old metal-poor clusters \citep[e.g.,][]{Forbes2018}, flat radial distribution for old metal-rich clusters are not unheard of \citep[e.g.,][]{Strader2012,Kartha2014}. Since the old metal-rich and young populations show significantly more scatter than the old metal-poor clusters, any initial gradient may have been erased by an accretion event not fully mixed yet.

\subsection{Kinematics}
\label{kinematics}

\begin{figure}
	\includegraphics[width=0.99\columnwidth]{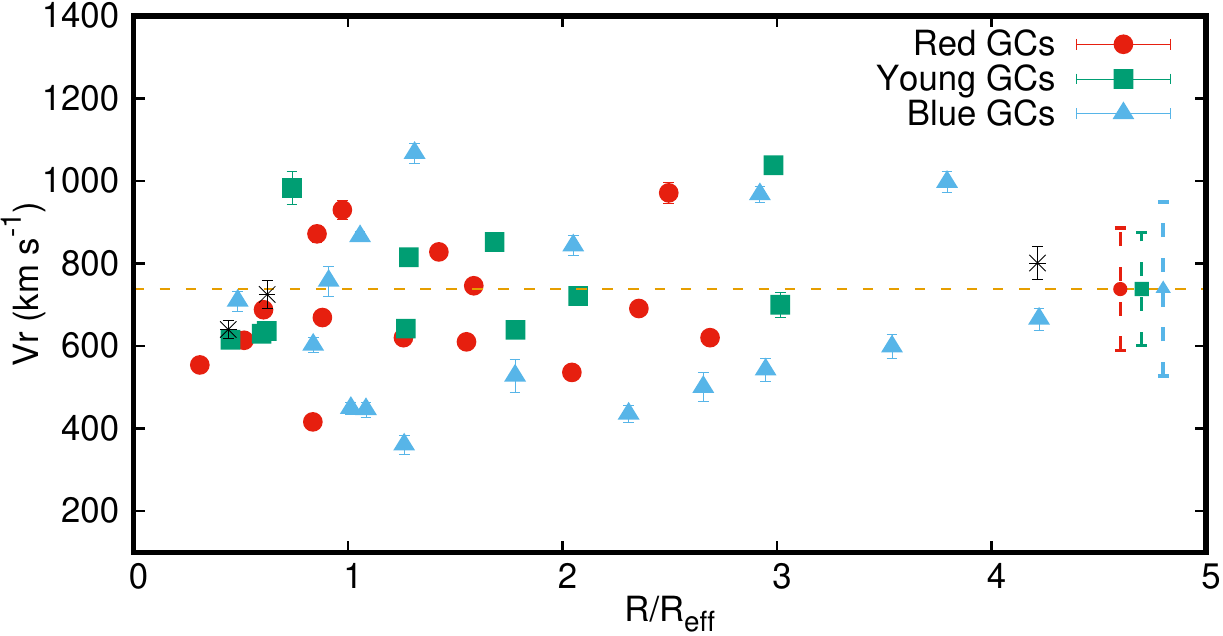}
    \caption{Radial velocity of the old metal-poor (blue triangles), old metal-rich (red circles) and young (green squares) GCs as a function of the galactocentric radius. Black asterisks indicate those objects whose stellar population parameters could not be determined. The standard deviation of radial velocity for the three GC subpopulations is indicated with dashed error bars. The dashed orange line traces the galaxy systemic velocity.}
    \label{fig:vel_dist}
\end{figure}

\begin{figure}
	\includegraphics[width=0.99\columnwidth]{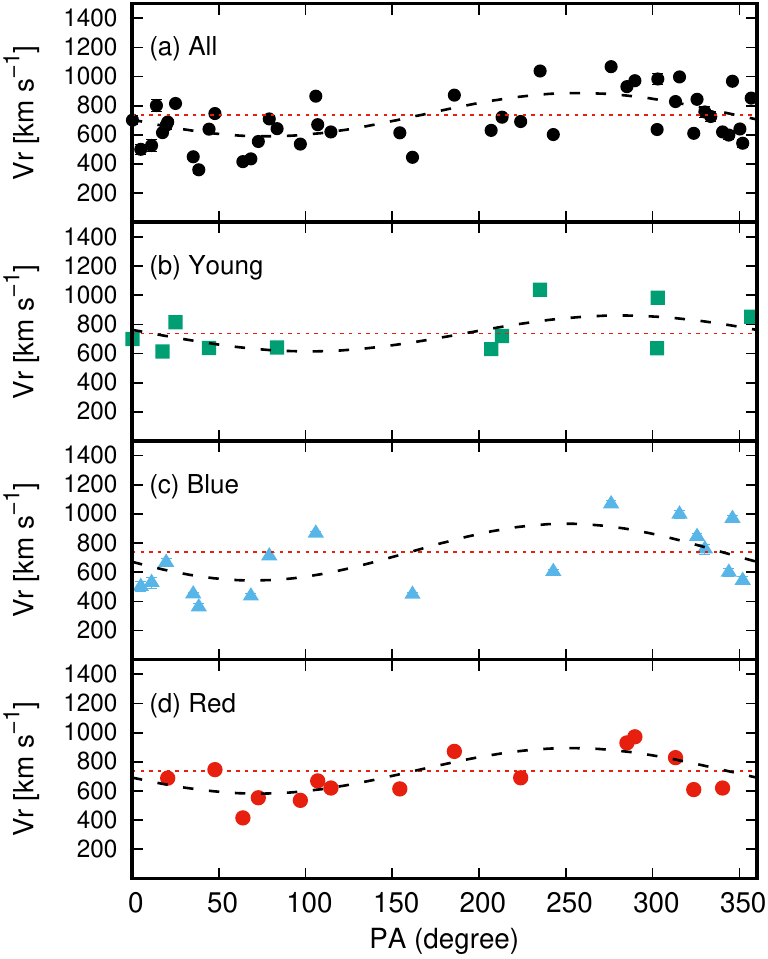}
    \caption{Radial velocities of all sample, the young, blue, and red GCs (panels a, b, c, and d, respectively) as a function of position angle (PA). The horizontal dotted lines indicate the radial velocity of the nucleus of NGC\,4382 obtained in this work ($738\pm6$ km\,s$^{-1}$), and the dashed lines the best fit.}
    \label{fig:kinemat_spec}
\end{figure}

Figure \ref{fig:vel_dist} shows the one-dimensional phase space diagram, obtained by plotting the variation of the GC radial velocity with radius, of the GC system of NGC\,4382. Old metal-poor and metal-rich (hereafter blue and red) GCs are shown as blue triangles and red circles, respectively, while young GCs are represented by green squares. Those objects whose stellar parameters could not be estimated (see Table \ref{Tab_vr}) are indicated with a black asterisk. The dashed orange line traces the galaxy systemic velocity. In Figure \ref{fig:vel_dist} it is possible to see that the objects nearest to the centre of NGC\,4382 ($R/R_\mathrm{eff}\lesssim0.7$), correspond, in their majority, to the red and young clusters. At all radii, red and young GCs have a radial velocity in a range of $\pm 200$ km\,s$^{-1}$ with respect to the systemic velocity of the galaxy, with the exception of few outliers. Blue GCs, instead, lie at a larger distance from the galaxy systemic velocity, $\Delta V_{r} \le 400 $. 
\citet{Cortesi2016} found a similar behaviour for the GC system of NGC\,1023, where red GCs are found to be rotating as the stars in the disk of the galaxy and blue GCs have the same kinematics as the halo component. 
To study this effect, we analyze the kinematics of the GC system as well as that of each subpopulation. Figure \ref{fig:kinemat_spec} shows the radial velocity of GCs as a function of the position angle (PA). At first glance, a sign of rotation is clearly evident in each of the subpopulations and in the system as a whole.
We calculate the rotation velocity of each system by fitting the following equation: 
\begin{equation}\label{eq:kinematics}
  V(\theta) = \Omega R\,\mathrm{sin}(\theta - \theta_0) + V_\mathrm{sys},
\end{equation}
where $V(\theta)$ is the velocity of each GC in the sample, $V_\mathrm{sys}$ the systemic velocity of the GC system, wherein this case we adopt the radial velocity of the galaxy ($738\pm6$ km\,s$^{-1}$; Table \ref{Tab_vr}), $\Omega R$ the amplitude of the projected rotation velocity, $\theta$ the azimuthal angle of each object relative to the galactic centre, and $\theta_0$ the orientation of the rotation angle. We perform a non-linear least-squares fit on the aforementioned samples. 
As in the work of K18,
we correct the amplitude values of the projected rotation velocity considering the inclination angle of the galaxy ($i=39$ degrees). 
Table \ref{Table_kinematics} lists the values obtained from the fits as well as the mean values of radial velocity and velocity dispersion for each sample.
The values obtained for the GC system are in very good agreement with those obtained by K18 ($\Omega R=148^{+67}_{-42}$ km\,s$^{-1}$, $\theta_0=161^{+25}_{-18}$ degrees) and for the inner ($R_\mathrm{gal}<6$ arcmin) GC system in K20 ($\Omega R=150^{+35}_{-37}$ km\,s$^{-1}$, $\theta_0=188^{+7}_{-13}$ degrees).
However, it is difficult to compare the values recovered for the subpopulations in this work and in K18, since they are differently defined. While, K18 separate the GCs into intermediate and old, in here we choose to study the red, blue and young GCs. We can, on the other hand, compare with the results obtained by K20 for the inner GCs. 
We estimate the $\Omega R/\sigma$ for the system and the subpopulations (young, blue and red), obtaining $0.85\pm0.19, 0.83\pm0.48, 0.89\pm0.34, 1.01\pm0.27$, respectively. K20 list a value of $\Omega R/\sigma_{r,corr}$ for the whole population and the red GC subpopulation, while it returns $\Omega R/\sigma_{r}$ for the blue subpopulation. We used the values of $\sigma_{r}$ from Table 5 of K20 to calculate $\Omega R/\sigma_{r}$ for the whole population in the bins that overlap with the sample in study here (i.e. till 5.9 and 5.3 arcmin for the whole population and the red subpopulation respectively). We obtain $\Omega R/\sigma_{r}=0.98^{+0.06}_{-0.11}$ for the whole GC system and $\Omega R/\sigma_{r}=1.46^{+0.17}_{-0.32}$ for the red subpopulation. For the blue subpopulation, K20 estimate $\Omega R/\sigma_{r}=0.41^{+0.64}_{-0.71}$ (they do not apply correction to the dispersion velocity since the rotation of blue GCs is negligible). These values are in agreement, within error bars, with the determination obtained in this work. Note the large errors on the $\Omega R/\sigma$ measured for the GC subpopulations, especially for the young and blue GCs. This results from both low number statistics in the fit, and, in the case of blue GCs, from the large spread of the radial velocities $V_\mathrm{r}$, with respect to the galaxy systemic velocity, which returns a high value of $\Omega R$ as well as of $\sigma$ (see Table \ref{Table_kinematics}). Such large errors for $\Omega R/\sigma$ for the blue subpopulation might also indicate a complex kinematic, resulting from multiple origins of the blue GCs \citep[see,][]{Cortesi2016, Shapiro2010}. In the following text, we discuss only the $\Omega R/\sigma$ value obtained for the whole population of GCs, which present the lowest error. 
According to simulation, for star like particles, the values of $\Omega R/\sigma$ recovered in this work for the the GC system is typical of minor merger with a mass ratio between 4.5:1 to 7:1 \citep{Bournaud2005}. On the other side, of the 71 S0-like merger remnants ($\Omega R/\sigma \simeq 1$) extracted by \citet{Tapia2017} from the GalMer database, 29 corresponds to minor encounters (7:1 to 20:1 mass ratio) and 42 to major ones, the majority (69\%) being the result of direct encounters and 31\% of retrograde encounters.
According to the results obtained in this work, the radial velocity dispersion of the GC system is similar to the velocity dispersion of the stars in NGC\,4382 \citep[$\sigma=179$ km\,s$^{-1}$ within 4-arcsec;][]{McDermid2006}. 
This is also found in NGC\,1023, where all the  GC subpopulations share a similar dispersion velocity, while the rotation velocity varies for the different subpopulations \citep{Cortesi2016}.
The old clusters (blue and red ones) have an identical rotation angle ($\sim$160 degrees), as in K18
and K20
for the blue GCs, while the rotation orientation of the young clusters is slightly greater ($\theta_0=191$ degrees) compared to the previous ones, but still within the errors, resembling the value found for red GCs in K20.
Interestingly, the position angle of the kinematic major axis of the GC system is perpendicular to the kinematic angle of the stars, $PA_\mathrm{kin} = 19.5 \pm 4.8$ degrees \citep{Krajnovic2011}, as determined using SAURON data. \citet{Diaz2016} modelled the SAURON 2D velocity field of this galaxy and found that it has a kinematically decoupled core that is counter-rotating relative to the rest of the galaxy. Yet, the field of view of SAURON covers the inner 30" of the galaxy, where few of the GCs lie. Moreover, \citet{Kormendy2009} found a variation of the position angle of the galaxy isophotes with radius, according to which the SAURON data might describe the bulge dominated area of the galaxy, and the red GCs would lie along the disk major axis. 
The comparison of the GC kinematics with long slit data covering larger radii along the major and minor axis will be studied in details in the following papers (Cortesi et al. in prep).

\begin{table}
\centering
\begin{tabular}{lccclc}
\multicolumn{6}{c}{}\\
\hline
\hline
\multicolumn{1}{c}{\textbf{Population}} &
\multicolumn{1}{c}{\textbf{$\langle V_r \rangle$}} &
\multicolumn{1}{c}{\textbf{$\sigma_r$}} &
\multicolumn{1}{c}{\textbf{$\Omega R$}} & 
\multicolumn{1}{c}{\textbf{$\Omega R^{*}$}} & 
\multicolumn{1}{c}{\textbf{$\theta_0$}} \\
\multicolumn{1}{c}{} &
\multicolumn{1}{c}{(km\,s$^{-1}$)} &
\multicolumn{1}{c}{(km\,s$^{-1}$)} & 
\multicolumn{1}{c}{(km\,s$^{-1}$)} &
\multicolumn{1}{c}{(km\,s$^{-1}$)} &
\multicolumn{1}{c}{(deg)} \\
\hline
All    & 698  & 175  & 148$\pm$33  &  235$\pm$52  & 167$\pm$11  \\

Blue   & 666  & 217  & 194$\pm$74  &  308$\pm$117 & 160$\pm$18  \\

Young  & 751  & 149  & 123$\pm$72  &  195$\pm$114 & 191$\pm$24  \\

Red    & 691  & 153  & 155$\pm$41  &  246$\pm$65  & 162$\pm$16  \\

\hline
\end{tabular}
\caption{Kinematic parameters of the GC system of NGC\,4382. $\langle V_r \rangle$: mean radial velocity, $\sigma_r$: velocity dispersion, $\Omega R$: amplitude of the projected rotation velocity, $\Omega R^{*}$: corrected-amplitude of the projected rotation velocity considering the inclination angle of the galaxy, $\theta_0$: orientation of the rotation angle.}
\label{Table_kinematics}
\end{table}


\subsection{Star Formation History of NGC4382 and its GCs}
\label{ngc4382}

As mentioned in Section \ref{spect_data}, we used the final 2D spectrum of the galaxy in order to extract 1D spectra along the slit at different galactocentric radii. To do this, we binned the spectrum in the spatial direction, applying increasing bin width towards the outer regions of the galaxy in order to obtain appropriate $S/N$ values ($S/N>20$ per pix). This $S/N$ value allows us to achieve reliable measurements of the kinematic and stellar populations parameters up to $\sim$75 arcsec (6.5 kpc) towards the south direction, and $\sim$25 arcsec (2.1 kpc) to the north direction, from the galactic centre.

\begin{figure}
	\includegraphics[width=0.99\columnwidth]{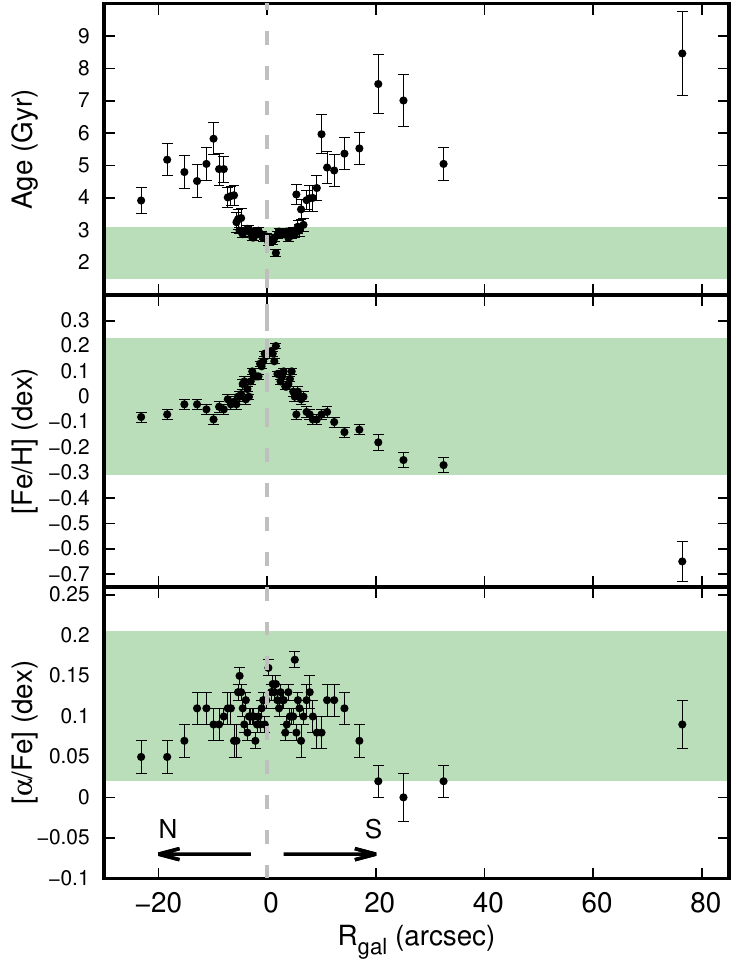}
    \caption{From top to bottom: luminosity-weighted age, metallicity and [$\alpha$/Fe] ratio as a function of the galactocentric radius obtained with ULySS. The vertical dashed line indicates the galactic centre. The green bands indicate the range in age, metallicity, and [$\alpha$/Fe] of the sample of young GCs. The black arrows indicate the north-south direction in which the slit is located (see Figure \ref{fig:NGC4382_fields}).}
    \label{fig:stellar_galaxy}
\end{figure}

As in the analysis carried out on the GCs, we used the ULySS code together with the SSP models resolved in [$\alpha$/Fe] built with PEGASE.HR model grid and the ELODIE3.2 library (see Section \ref{full_spectral}), on each extracted spectrum. 
In this paper, we will not focus on the kinematic analysis of the galaxy, which will be addressed in details in Cortesi et al. (in prep). 
On the other hand, the analysis of the stellar populations of NGC\,4382 was carried out via full spectral fitting, using a multiplicative polynomial of order 10 together with the {\sc{clean}} option of ULySS. 
Figure \ref{fig:stellar_galaxy} shows the luminosity-weighted SSP equivalent ages, metallicities and [$\alpha$/Fe] values as a function of the galactocentric radius. As can be seen in the figure, the central region of the galaxy ($R_\mathrm{gal}<10$ arcsec; 0.87 kpc) has a young population with a luminosity-weighted mean age of $\sim$2.7 Gyr. From this radius, the age begins to increase up to $\sim7$ Gyr at $R_\mathrm{gal}=20$ arcsec (1.7 kpc) towards the southern region of the galaxy, while towards the north direction it reaches an age of $\sim5$ Gyr at $R_\mathrm{gal}=15$ arcsec (1.3 kpc). After this radius, the age values obtained show a slight decrease. However, this should be taken with caution due to the low number of points.

On the other hand, the radial metallicity distribution shows higher values towards the galactic centre than in the outskirts. The metallicity increases from $-0.1$ dex at $R_\mathrm{gal}=10$ arcsec up to $+0.2$ dex in the center. Finally, the [$\alpha$/Fe] values remain relatively constant ($\sim$0.12 dex) at $R_\mathrm{gal}<18$ arcsec (1.6 kpc). This last value is in good agreement with that obtained by \citet{McDermid2006} ([$\alpha$/Fe]=0.12$\pm$0.06 dex within $R=4$ arcsec).

When comparing these values obtained for the galaxy with those of the spectroscopic sample of GCs, we observed a good agreement between the different stellar population parameters of NGC\,4382 at $R_\mathrm{gal}<10$ arcsec, and the sample of young GCs  (green bands in Figure \ref{fig:stellar_galaxy}; $2.2\pm0.9$ Gyr, $\mathrm{[Fe/H]}=-0.05\pm0.28$ dex, $\mathrm{[\alpha/Fe]}=0.11\pm0.10$ dex). This similarity suggests that NGC\,4382 merged with a gas-rich object about 2.7 Gyr ago, which gave rise to this subpopulation of younger GCs as well as the galaxy's young diffuse stellar population. 
For their part, the ancient ages and low metallicities of the blue and red GC subpopulations obtained in this work probe that these GCs formed earlier in the lifetime of the galaxy.

\section{Summary and Conclusions}
\label{conclusions}

In order to provide a clearer picture of the complex GC system associated with the galaxy NGC\,4382, we present a detailed photometric and spectroscopic study of it using Gemini/GMOS data.

From the photometric analysis, using the Gaussian Mixture Model (GMM) code in the colour plane $(g'-i')_0$ versus $(g'-r')_0$, we identified 4 different groups of clusters, the typical blue ($(g'-i')_0 \sim 0.77$ mag; group A) and red ($(g'-i')_0 \sim 1.07$ mag; group C) subpopulations, a group with intermediate colours to the previous ones ($(g'-i')_0 \sim 0.92$ mag, group B), and a fourth group with redder colours ($(g'-i')_0 \sim 1.21$ mag, group D). 

These results indicate a partial agreement with previous work by K19, in the complex colour distribution of the GC system within the inner 4 arcmin (20.9 kpc). However, our analysis in the colour plane shows that the red GCs presents mean colours typical for this type of subpopulations, unlike that obtained by K19 where this subpopulation is $\sim$0.1 mag bluer in comparison with other Virgo early-type galaxies. We attribute this difference to the mixing of an underlying GC subpopulation with intermediate colours, untraceable without a careful analysis of a broader sample of GCs with individual spectral follow-up, as we show in this paper.

When analyzing the colour distribution as a function of the projected galactocentric radius up to $R_\mathrm{gal}<5$ arcmin ($R_\mathrm{gal}<26.1$ kpc) of these 4 groups of GCs, it is observed that the group with intermediate colours shows two clumps located at 1 arcmin (5.2 kpc) and 2.1 arcmin (10.9 kpc). In particular, some objects that make up the second clump are found in regions where different stellar shells are observed (see Figure \ref{fig:distrib_espacial}). This tentatively suggests that the GC group with intermediate colours is associated with the same merger event responsible for the shells observed in this system. Furthermore, this region coincides with the region where an unusual "extra-halo" is reported in the surface brightness profile of the galaxy \citep{Kormendy2009}. However, a future spectroscopic analysis will be necessary to confirm or rule out the possible association between these structures and the GCs. 

For its part, the group of objects with colours redder than the typical red GCs make up approximately 8\% of the photometric sample of GC candidates and are located near the galactic centre at $R_\mathrm{gal}<2$ arcmin (10.4 kpc). The presence of this group could be associated with those objects called diffuse star clusters (DSC), of which the galaxy has shown to present an excess \citep{Peng2006b,Liu2016}. Faint DSCs could be the objects identified as {\it faint fuzzies} (FF) by \citet{Brodie2002}, which have been suggested to be connected to the formation of lenticular galaxies \citep{Burkert2005}. Interestingly, \citet{Larsen2001} found that not all lenticular galaxies present FFs, as in the case of NGC\,3115. The presence of FFs could be related to the formation mechanisms responsible for the origin of lenticular galaxies. Maybe, FFs are present in S0s that are faded spirals, as galaxies belonging to cluster and groups, which is the case of NGC\,4382, NGC\,1023 and NGC\,3384 \citep{Brodie2002,Cortesi2016}. While, isolated S0s galaxies, as NGC\,3115, which possibly formed through disk instability and accretion will not present such populations \citep{Burkert2005}.   

From the spectroscopic analysis, we determined the radial velocity of 53 objects
confirming 47 GCs associated with NGC\,4382. The mean radial velocity and the velocity dispersion of this data set results in $\mu=674\pm35$ km\,s$^{-1}$ and $\sigma=158\pm36$ km\,s$^{-1}$, with the value of the mean radial velocity being slightly less than the systemic velocity of the galaxy \citep[$V_\mathrm{sys}=729\pm2$ km\,s$^{-1}$;][]{Smith2000}. 

We analyzed the luminosity-weighted stellar population parameters in the data set (age, metallicity, and $\alpha$-element abundance) using the full spectral fitting technique through the ULySS software together with the PEGASE-HR models. This spectroscopic analysis allowed us to confirm 3 of the 4 groups identified by GMM on the photometric data, the typical blue and red GCs (groups A and C), and the group with intermediate colours (group B). On the other hand, given that only one object in the spectroscopic sample would be associated with group D, we cannot confirm the presence of this group. 
The different characteristics of each GC subpopulation are detailed below. 

\begin{itemize}
  \item The blue GC subpopulation has a mean radial velocity of 666 km\,s$^{-1}$ with a radial velocity dispersion of 217 km\,s$^{-1}$. This group presents rotation with a corrected rotation amplitude of $\Omega R^{*} = 308\pm117$ km\,s$^{-1}$ and a rotation axis of 160$\pm$18 degrees. On the other hand, the mean values of its stellar population parameters result in $10.4\pm2.8$ Gyr, [Fe/H] =$-1.48\pm0.18$ dex, and [$\alpha$/Fe] = $0.24\pm0.16$ dex, which are typical values for this type of subpopulation. Furthermore, this group clearly shows a negative metallicity gradient ([Fe/H] =($-1.70_{-0.07}^{+0.07})-(0.27_{-0.06}^{+0.06})\,log (R/R_\mathrm{eff})$) as we move away from the galactic centre. This characteristic would indicate that the origin of the blue GCs is due to early dissipative events towards the inner regions of the galaxy with the subsequent accretion of GCs as a result of mergers with low mass galaxies \citep[e.g.,][]{Forbes2011}.

  The optical/near-infrared study of \citet{Chies-Santos2011} presents evidence for this galaxy hosting younger GCs if compared to GC systems with typically old ages ($\geqslant 10$ Gyr) such as NGC\,4486 and NGC\,4649. Such study reports tantalizing evidence that bluer GCs could be the clusters driving this age trend. From the 20 objects in common between the present work and that of \citet{Chies-Santos2011}, we find that 3 of these objects have younger ages ($\sim$ 2 Gyr) and metallicities from $-0.45$ to $+0.19$ dex and show intermediate colours in our photometric analysis.

  \item For its part, the red GC subpopulation has a mean radial velocity of 691 km\,s$^{-1}$, being closer to the value of the systemic velocity of the galaxy, with a radial velocity dispersion of 153 km\,s$^{-1}$. This group shows a clear rotation perpendicular to the stellar rotation axis \citep{Krajnovic2011,Ko2018,Ko2020}, with a corrected rotation amplitude of $\Omega R^{*} = 246\pm65$ km\,s$^{-1}$ and a similar rotation axis to that of the blue ones of 162$\pm$16 degrees. The position angle of the stellar kinematics is retrieved from IFU data within a radius of 20 arcsec which might cover the bulge dominated part of the galaxy. The galaxy isophotal profile varies with radius \citep{Kormendy2009}, suggesting that the red GCs might be rotating along the disk major axis, as in NGC\,1023 \citep{Cortesi2016}. Regarding its stellar population parameters, the mean age for the reds is $12.1\pm2.3$ Gyr, being slightly older compared to the blue GCs, although within the uncertainties. On the other hand, its metallicity and $\alpha$-element enhancement result in [Fe/H]=$-0.64\pm0.26$ dex and [$\alpha$/Fe]=$0.24\pm0.10$ dex, being typical values for this subpopulation. 
  
  \item As in K18, a group of young GCs stands out in the system of NGC\,4382. When analyzing the stellar population parameters of this group, a mean age of $2.2\pm0.9$ Gyr, metallicity of [Fe/H] = $-0.05\pm0.28$ dex and [$\alpha$/Fe] = $0.11\pm0.10$ dex, is obtained. These young GCs present a mean radial velocity of 751 km\,s$^{-1}$ with a radial velocity dispersion of 149 km\,s$^{-1}$, showing a $\Omega R^{*} = 195\pm114$ km\,s$^{-1}$ with a rotation angle (191$\pm$24 degrees) slightly different from the aforementioned subpopulations. On the other hand, the photometric colours of this young subpopulation are in good agreement with group B identified in the photometric analysis carried out in this work. 
  
  \item Finally, it was not possible to spectroscopically confirm the photometric group D, which would be composed of the so-called diffuse star clusters \citep{Peng2006b}. Although one object in our spectroscopic sample (GC(166)) presents colours characteristic of this group, both its brightness ($g'_0=22.252$ mag) and metallicity ([Fe/H]=$-0.54\pm0.06$ dex) values are not typical for this type of objects. This is why GC(166) is probably a typical red GC. 
  
\end{itemize}

We also analyze the stellar populations of NGC\,4382 itself as a function of the galactocentric radius from long-slit data-oriented along the north-south direction of the galaxy. Towards the central region ($R_\mathrm{gal}<10$ arcsec; 0.87 kpc), NGC\,4382 presents a luminosity-weighted mean age of $\sim$2.7 Gyr, and a metallicity increasing from [Fe/H]=$-0.1$ dex to $+0.2$ dex in the mentioned region. These values are in good agreement with those obtained on the young GC population, which could suggest a common origin, as a consequence of a recent (minor) merger that the galaxy experienced. Outside $R_\mathrm{gal}=10$ arcsec, the age of the galaxy begins to increase until reaching 7 Gyr at $R_\mathrm{gal}=20$ arcsec (1.73 kpc), while the metallicity shows a negative gradient towards the outer regions. Finally, the parameter [$\alpha$/Fe] has a relatively constant value of $+0.12$ dex at $R_\mathrm{gal}<18$ arcsec (1.56 kpc). 

Combining all these findings, we can imagine a formation scenario for this galaxy where old metal-poor blue GCs are associated with the galaxy halo, old metal-rich red GCs were formed through the merger of clumps at high redshifts ($z \simeq 2$) \citep{Shapiro2010,Cortesi2016,Saha2018} and the young GCs have originated in a recent encounters with neighbouring galaxies, as well as the inner stellar population. In a way, NGC\,4382 is consistent with being a lenticular galaxy in the process of turning into an elliptical galaxy, through  mergers and encounters that slowly disrupt the galaxy disk.

\section*{Acknowledgments}
We thank the anonymous referee for his/her constructive comments.
A.C.S. acknowledges funding from CNPq and the Rio Grande do Sul Research Foundation (FAPERGS) through grants CNPq-403580/2016-1, CNPq-11153/2018-6, PqG/FAPERGS-17/2551-0001, FAPERGS/CAPES 19/2551-0000696-9, L’Or\'eal UNESCO ABC \emph{Para Mulheres na Ci\^encia} and the Chinese Academy of Sciences (CAS) President's International Fellowship Initiative (PIFI) through grant E085201009.
E.J.J. acknowledges support from FONDECYT Iniciaci\'on en investigaci\'on Project 11200263.
A.C. acknowledges the support of CAPES and of the Laboratório Nacional de Astrofísica (LNA, Brazil), in particular of Dr Eder Martioli.
This work was funded with grants from Consejo Nacional de Investigaciones Cientificas y Tecnicas de la Republica Argentina, and Universidad Nacional de La Plata (Argentina). 
Based on observations obtained at the international Gemini Observatory, a program of NSF’s NOIRLab, which is managed by the Association of Universities for Research in Astronomy (AURA) under a cooperative agreement with the National Science Foundation on behalf of the Gemini Observatory partnership: the National Science Foundation (United States), National Research Council (Canada), Agencia Nacional de Investigaci\'{o}n y Desarrollo (Chile), Ministerio de Ciencia, Tecnolog\'{i}a e Innovaci\'{o}n (Argentina), Minist\'{e}rio da Ci\^{e}ncia, Tecnologia, Inova\c{c}\~{o}es e Comunica\c{c}\~{o}es (Brazil), and Korea Astronomy and Space Science Institute (Republic of Korea).
The Gemini program ID are GN-2006A-Q-81, GN-2009A-Q-102, GN-2014A-Q-35, GN-2016A-Q-62, GN-2015A-Q-207 and GN-2019A-Q-903.

\section*{Data Availability}
The data underlying this article will be shared on reasonable request to the corresponding author.



\bibliographystyle{mnras}
\bibliography{NGC4382.bib}








\bsp	
\label{lastpage}
\end{document}